\RequirePackage[table]{xcolor}
\documentclass[table]{article}
\usepackage{lmodern}
\usepackage{amssymb,amsmath}
\usepackage{ifxetex,ifluatex}
\usepackage{algorithm}
\usepackage{algpseudocode}
\ifnum 0\ifxetex 1\fi\ifluatex 1\fi=0 % if pdftex
  \usepackage[T1]{fontenc}
  \usepackage{textcomp} % provide euro and other symbols
  \usepackage[utf8]{inputenc}
\else % if luatex or xelatex
  \ifxetex
    \usepackage{mathspec}
  \else
    \usepackage{fontspec}
  \fi
  \defaultfontfeatures{Ligatures=TeX,Scale=MatchLowercase}
\fi
\IfFileExists{upquote.sty}{\usepackage{upquote}}{}
\IfFileExists{microtype.sty}{%
\usepackage{microtype}
\UseMicrotypeSet[protrusion]{basicmath} % disable protrusion for tt fonts
}{}
\usepackage[margin=1in]{geometry}
\usepackage{hyperref}
\hypersetup{unicode=true,
            pdftitle={A Tutorial on Time-Dependent Cohort State-Transition Models in R using a Cost-Effectiveness Analysis Example},
            pdfauthor={Fernando Alarid-Escudero, PhD; Eline Krijkamp, MSc; Eva A. Enns, PhD; Alan Yang, MSc; M. G. Myriam Hunink, PhD\^{}\textbackslash{}dagger; Petros Pechlivanoglou, PhD; Hawre Jalal, MD, PhD},
            pdfkeywords={Markov models, state-transition models, decision models, Tutorial, R},
            pdfborder={0 0 0},
            breaklinks=true}
\usepackage{color}
\usepackage{fancyvrb}

\DefineVerbatimEnvironment{Highlighting}{Verbatim}{commandchars=\\\{\}}
\usepackage{framed}
\definecolor{shadecolor}{RGB}{248,248,248}
\newenvironment{Shaded}{\begin{snugshade}}{\end{snugshade}}

\newcommand{\DecValTok}[1]{\textcolor[rgb]{0.00,0.00,0.81}{#1}}

\newcommand{\FloatTok}[1]{\textcolor[rgb]{0.00,0.00,0.81}{#1}}

\newcommand{\SpecialCharTok}[1]{\textcolor[rgb]{0.00,0.00,0.00}{#1}}
\newcommand{\StringTok}[1]{\textcolor[rgb]{0.31,0.60,0.02}{#1}}

\newcommand{\CommentTok}[1]{\textcolor[rgb]{0.56,0.35,0.01}{\textit{#1}}}
\newcommand{\DocumentationTok}[1]{\textcolor[rgb]{0.56,0.35,0.01}{\textbf{\textit{#1}}}}

\newcommand{\OtherTok}[1]{\textcolor[rgb]{0.56,0.35,0.01}{#1}}
\newcommand{\FunctionTok}[1]{\textcolor[rgb]{0.00,0.00,0.00}{#1}}

\newcommand{\ControlFlowTok}[1]{\textcolor[rgb]{0.13,0.29,0.53}{\textbf{#1}}}

\newcommand{\AttributeTok}[1]{\textcolor[rgb]{0.77,0.63,0.00}{#1}}

\newcommand{\NormalTok}[1]{#1}
\usepackage{longtable,booktabs}
\usepackage{calc} % for calculating minipage widths
\usepackage{graphicx,grffile}
\makeatletter
\def\maxwidth{\ifdim\Gin@nat@width>\linewidth\linewidth\else\Gin@nat@width\fi}
\def\maxheight{\ifdim\Gin@nat@height>\textheight\textheight\else\Gin@nat@height\fi}
\makeatother
\setkeys{Gin}{width=\maxwidth,height=\maxheight,keepaspectratio}
\IfFileExists{parskip.sty}{%
\usepackage{parskip}
}{% else
\setlength{\parindent}{0pt}
\setlength{\parskip}{6pt plus 2pt minus 1pt}
}
\ifx\paragraph\undefined\else
\let\oldparagraph\paragraph
\renewcommand{\paragraph}[1]{\oldparagraph{#1}\mbox{}}
\fi
\ifx\subparagraph\undefined\else
\let\oldsubparagraph\subparagraph
\renewcommand{\subparagraph}[1]{\oldsubparagraph{#1}\mbox{}}
\fi

\let\rmarkdownfootnote\footnote%
\def\footnote{\protect\rmarkdownfootnote}

\usepackage{titling}

\newlength{\cslhangindent}
\newlength{\csllabelwidth}
\newenvironment{CSLReferences}[2] % #1 hanging-ident, #2 entry spacing
 {% don't indent paragraphs
  \setlength{\parindent}{0pt}
  \ifodd #1 \everypar{\setlength{\hangindent}{\cslhangindent}}\ignorespaces\fi
  \ifnum #2 > 0
  \setlength{\parskip}{#2\baselineskip}
  \fi
 }%
 {}
\usepackage{calc}

\newcommand{\CSLLeftMargin}[1]{\parbox[t]{\csllabelwidth}{#1}}
\newcommand{\CSLRightInline}[1]{\parbox[t]{\linewidth - \csllabelwidth}{#1}\break}

  \title{A Tutorial on Time-Dependent Cohort State-Transition Models in R using a Cost-Effectiveness Analysis Example}
    \pretitle{\vspace{\droptitle}\centering\huge}
  \posttitle{\par}
    \author{Fernando Alarid-Escudero, PhD\footnote{Division of Public Administration, Center for
  Research and Teaching in Economics (CIDE), Aguascalientes,
  AGS, Mexico} \\ Eline Krijkamp, MSc\footnote{Department of Epidemiology and Department of Radiology, Erasmus
  University Medical Center, Rotterdam, The Netherlands} \\ Eva A. Enns, PhD\footnote{Division of Health Policy and Management,
  University of Minnesota School of Public Health, Minneapolis, MN, USA} \\ Alan Yang, MSc\footnote{The Hospital for Sick Children,
  Toronto, Ontario, Canada} \\  M.G. Myriam Hunink, PhD\(^\dagger\)\footnote{Center for Health Decision
  Sciences, Harvard T.H. Chan School of Public Health, Boston, USA} \\ Petros Pechlivanoglou, PhD\footnote{The Hospital for Sick Children,
  Toronto and University of Toronto, Toronto, Ontario, Canada} \\ Hawre Jalal, MD, PhD\footnote{University of Ottawa, Ottawa, ON,
  Canada}}
    \preauthor{\centering\large\emph}
  \postauthor{\par}
      \predate{\centering\large\emph}
  \postdate{\par}
    \date{2022-06-01}

\usepackage{booktabs}
\usepackage{longtable}
\usepackage{array}
\usepackage{multirow}
\usepackage{wrapfig}
\usepackage{float}
\usepackage{colortbl}
\usepackage{pdflscape}
\usepackage{tabu}
\usepackage{threeparttable}
\usepackage{threeparttablex}
\usepackage[normalem]{ulem}
\usepackage{makecell}

\usepackage{amsmath}
\usepackage{float}
\usepackage{setspace}

\begin{document}
\maketitle
\begin{abstract}
In an introductory tutorial, we illustrated building cohort state-transition models (cSTMs) in R, where the state transitions probabilities were constant over time. However, in practice, many cSTMs require transitions, rewards, or both to vary over time (time-dependent). This tutorial illustrates adding two types of time-dependency using a previously published cost-effectiveness analysis of multiple strategies as an example. The first is simulation-time dependence, which allows for the transition probabilities to vary as a function of time as measured since the start of the simulation (e.g., varying probability of death as the cohort ages). The second is state-residence time dependence, allowing for history by tracking the time spent in any particular health state using tunnel states. We use these time-dependent cSTMs to conduct cost-effectiveness and probabilistic sensitivity analyses. We also obtain various epidemiological outcomes of interest from the outputs generated from the cSTM, such as survival probability and disease prevalence, often used for model calibration and validation. We present the mathematical notation first, followed by the R code to execute the calculations. The full R code is provided in a public code repository for broader implementation.

\end{abstract}

%{
%\setcounter{tocdepth}{2}
%\tableofcontents
%}
\hypertarget{introduction}{%
\section{Introduction}\label{introduction}}

Cohort state-transition models (cSTMs), commonly known as Markov models, are decision models that simulate disease dynamics over time. cSTMs are widely used in health decision sciences to evaluate various health policies and clinical strategies, such as screening and surveillance programs,\textsuperscript{\protect\hyperlink{ref-Suijkerbuijk2018}{1},\protect\hyperlink{ref-Sathianathen2018a}{2}} diagnostic procedures,\textsuperscript{\protect\hyperlink{ref-Lu2018b}{3}} disease management programs,\textsuperscript{\protect\hyperlink{ref-Djatche2018}{4}} and interventions.\textsuperscript{\protect\hyperlink{ref-Pershing2014}{5},\protect\hyperlink{ref-Smith-Spangler2010}{6}}. The simplest cSTMs are time-independent (time-homogeneous), meaning that the transition probabilities and other model parameters remain fixed over the simulated time horizon. The implementation of time-independent cSTMs is described in an introductory tutorial.\textsuperscript{\protect\hyperlink{ref-Alarid-Escudero2022b}{7}} However, a time-independent cSTM is limited in many applications, because key parameters may vary over time. For example, background mortality changes as the cohort ages, the risk of cancer recurrence might change as a function of time since diagnosis, or the incidence of vector-borne diseases may have temporal or seasonal trends. The costs and utility of residing in a particular health state might also vary over time. For example, cancer-related health care costs and utilities might depend on whether a patient is in their first year of remission or their fifth. Similarly, a person's last year of life is often the most expensive in health care spending. Therefore, capturing time-varying dynamics is often essential in realistic models.

This tutorial expands the cSTM framework described in the introductory time-independent cSTM tutorial by allowing time dependence on transition probabilities, costs, and utilities. We distinguish between two types of time-dependency that require different approaches: (1) simulation-time dependence, which represents dependence on the time since the start of the simulation, and (2) state-residence time dependence, representing dependence on the time spent in a health state. A common example of simulation-time dependence is age-specific background mortality. Since all members of the cohort in the simulation age together, we can implement this dependency by changing the transition to death as the simulation progresses.\textsuperscript{\protect\hyperlink{ref-Snowsill2019}{8}} Similarly, in a model simulating a cohort starting from disease diagnosis, any dependence on time since diagnosis can be implemented as dependent on the time since the simulation starts. Simulation-time dependence can also be used to incorporate seasonal or temporal variation in disease incidence, where the risk of developing a disease can vary based on the current season or year in the simulation.

Unlike simulation-time dependence, state-residence time dependence captures time dependence on events that members of the cohort could experience at different times. For example, members of a cohort of healthy individuals may experience disease onset at different times. Thus, parameters that depend on the time since disease onset cannot be implemented based on simulation time; instead, we need to track cohort members \emph{from their disease onset time}. We implement state-residence time dependence by expanding the model state space to capture the history of the disease since disease onset. For example, instead of a single ``Sick'' state, individuals would transition first to the ``Sick - cycle 1'' state at disease onset, then ``Sick - cycle 2'' at the next cycle, and so on. Each intermediate ``Sick'' state can have different transition probabilities (e.g., mortality, risk of complications, etc.), costs, or utilities. We can also use this state expansion approach to model simulation-time dependence. However, as we see below, substantial state expansion can be cumbersome and potentially computationally expensive. For simplicity, modelers should always consider simulation-time dependence first. Besides describing the two forms of time dependence, we illustrate the concept and implementation of transition rewards, which capture one-time event-driven outcomes (e.g., costs or utilities) that individuals experience when transitioning between two states.\textsuperscript{\protect\hyperlink{ref-Krijkamp2019}{9}} For example, a higher cost immediately at disease onset due to diagnostic tests or the one-time cost of transition to the dead state incurred from end-of-life interventions or management.

This tutorial describes how to incorporate simulation-time and state-residence time dependence in a cSTM in R\textsuperscript{\protect\hyperlink{ref-Jalal2017b}{10}} using a cost-effectiveness analysis (CEA) example. We illustrate the calculation of various epidemiological outcomes beyond the simple cohort trace, including survival, life expectancy, and prevalence. Such epidemiological outcomes can be informative, but they are also used to calibrate and validate the model against observed real-world data.

Readers can find the R code for this tutorial's analysis and graphs in the accompanying GitHub repository (\url{https://github.com/DARTH-git/cohort-modeling-tutorial-timedep}). We encourage readers first to review the basics of decision modeling\textsuperscript{\protect\hyperlink{ref-Hunink2014}{11}} and the introductory tutorial on time-independent cSTMs in R.\textsuperscript{\protect\hyperlink{ref-Alarid-Escudero2022b}{7}}

\hypertarget{simulation-time-dependence}{%
\section{Simulation-time dependence}\label{simulation-time-dependence}}

Simulation-time dependence represents the time since the model starts and can be represented by defining the transition probability matrix as a function of time, \(P_t\),
\[
  P_t = 
  \begin{bmatrix}
    p_{[1,1,t]} & p_{[1,2,t]} & \cdots & p_{[1,n_S,t]} \\
    p_{[2,1,t]} & p_{[2,2,t]} & \cdots & p_{[2,n_S,t]} \\
    \vdots    & \vdots  & \ddots & \vdots   \\
    p_{[n_S,1,t]} & p_{[n_S,2,t]} & \cdots & p_{[n_S,n_S,t]} \\
  \end{bmatrix},
\]
where \(p_{[i,j,t]}\) is the transition probability of moving from state \(i\) to state \(j\) in time \(t\), \(\{i,j\} = 1,\ldots, n_S\), \(t = 0,\ldots,n_T\), \(n_S\) is the number of health states of the model, and \(n_T\) is the number of cycles that represent total simulation time. We implement this dependence in R with a three-dimensional transition probability array, \(\mathbf{P}\), coded in R as \texttt{a\_P}, where the third dimension represents the time since the start of the simulation. Note that in each cycle \(t\) all rows of the transition probability matrix must sum to one, \(\sum_{j=1}^{n_S}{p_{[i,j,t]}} = 1\) for all \(i = 1,\ldots,n_S\) and \(t = 0,\ldots, n_T\).

Next, we specify the initial distribution of the cohort at \(t = 0\). We then define \(\mathbf{m}_{0}\) as the row vector that captures the distribution of the cohort among the states at \(t = 0\) (i.e., the initial state vector). As illustrated in the introductory tutorial, we iteratively compute the cohort distribution across the health states in each cycle using the transition probability matrix and the cohort's distribution at the prior cycle. The row state vector at the next cycle \(t+1\) (\(\mathbf{m}_{t+1}\)) is then calculated as the matrix product of the row state vector at cycle \(t\), \(\mathbf{m}_{t}\), and the transition probability matrix that the cohort faces in cycle \(t\), \(P_t\),

\begin{equation}
  \label{eq:time-dep-matrix-mult}
  \mathbf{m}_{t+1} = \mathbf{m}_{t} P_t \qquad\text{ for }\qquad t = 0,\ldots, (n_T - 1).
\end{equation}

Equation \eqref{eq:time-dep-matrix-mult} is iteratively evaluated until \(t = n_T\).

The cohort trace matrix, \(M\), is a matrix of dimensions \((n_T+1) \times n_S\) where each row is a state vector \((-\mathbf{m}_{t}-)\), such that

\[
  M = 
  \begin{bmatrix}
    - \mathbf{m}_0 -  \\
    - \mathbf{m}_1 -  \\
     \vdots \\
    - \mathbf{m}_{n_T} -  
  \end{bmatrix}. 
\]

\(M\) stores cohort's distribution across states over time. Thus, \(M\) can be used to compute various epidemiological outcomes, such as prevalence and survival probability over time, and economic outcomes, such as cumulative resource use and costs.

\hypertarget{time-dependence-on-state-residence}{%
\section{Time dependence on state-residence}\label{time-dependence-on-state-residence}}

The implementation of state-residence time dependence is more involved than simulation-time dependence. To account for state-residence time dependence, we expand the state of interest with as many transient states as the number of cycles required to track the history of the cohort in that state. For example, if a transition rate decreases over the first three cycles spent in a state and then is constant from the fourth cycle onward, four transient states would be needed for that state. These transient states are often called \emph{tunnel} states, where the cohort resides in each transient state for exactly one cycle, after which they either exit the tunnel or transition to the next tunnel state. The total number of states for a cSTM with tunnels is \(n_{S_\text{tunnels}} = n_S + n_{\text{tunnels}} - 1\), where \(n_{\text{tunnels}}\) is the total number of times a health state needs to be expanded to capture the relevant residence-time dependence. We subtract one because the original state is replaced by the \(n_{\text{tunnels}}\) tunnel states. The transition probability matrix also needs to be expanded to incorporate these additional transient states, resulting in a transition probability matrix of dimensions \(n_{S_\text{tunnels}} \times n_{S_\text{tunnels}}\). If the transition probabilities are also dependent on simulation time, as described in the previous section, we need to expand to a 3-dimensional transition array, with dimensions \(n_{S_\text{tunnels}} \times n_{S_\text{tunnels}} \times n_T\). 

The algorithm below (Algorithm \ref{alg:fill-tunnels}) describes populating the transition probability array for state-residence time-dependent cSTMs. 

\begin{algorithm}
\caption{Populate transitions from, to, and within state expanded with tunnels.}\label{alg:fill-tunnels}
\begin{algorithmic}
  \Ensure $P_t$ is the transition probability matrix for cycle $t$, $P_t[i, j]$ is the entry corresponding to the transition probability from state $i$ to state $j$ at time $t$, parameterized as $p_{[i,j,t]}$.
  \Ensure $r$ is the state expanded as a tunnel with $n_{\text{tunnels}}$ tunnel states and $r_{\tau}$ is the tunnel state corresponding to state-residence cycle $\tau$, where $\tau=1,\ldots,n_{\text{tunnels}}$.
  \For{For all $i$ and $j$ states different from $r$}
    \State $P_t[i, j] = p_{[i,j,t]}$ \Comment{This is similar to the simulation-time dependence case}
  \EndFor
  \For{States $i$ that can transition to the $r$ state}
    \State $P_t[i, r_{1}] = p_{[i,r_{1},t]}$
  \EndFor
  \For{States $i$ that can transition from any $r_{\tau}$ state}
    \For{$\tau = 1$ to $n_{\text{tunnels}}$}
      \State $P_t[r_{\tau}, i] = p_{[r_{\tau}, i,\tau]}$
      \If{$\tau = n_{\text{tunnels}}$}
        \State $P_t[r_{\tau}, r_{\tau}] = p_{[r_{\tau}, r_{\tau},\tau]}$
      \Else
        \State $P_t[r_{\tau}, r_{\tau+1}] = p_{[r_{\tau}, r_{\tau+1},\tau]}$
      \EndIf
    \EndFor
  \EndFor
\end{algorithmic}
\end{algorithm}

The transition probability array with tunnels is populated similarly to the simulation-time dependence case. However, we need a different approach for the state expanded with tunnel states and any state(s) with transitions to or from the tunnel. Filling in the transition probabilities for these states requires iterating through the tunnel states. There is typically some probability that the cohort remains in that state in the last tunnel state, like for other non-transient states. Here we only discuss models with a single tunnel. The same principles can be applied for adding more tunnels as needed. 

Table \ref{tab:Timedep-cSTM-components-table} describes the core components of time-dependent cSTMs and their suggested R code names. The Supplementary Material contains a more detailed description of these components with their variable types, including their R data structure and dimensions.

\begin{longtable}[]{@{}
  >{\raggedright\arraybackslash}p{(\columnwidth - 6\tabcolsep) * \real{0.19}}
  >{\raggedright\arraybackslash}p{(\columnwidth - 6\tabcolsep) * \real{0.57}}
  >{\centering\arraybackslash}p{(\columnwidth - 6\tabcolsep) * \real{0.17}}
  >{\raggedright\arraybackslash}p{(\columnwidth - 6\tabcolsep) * \real{0.06}}@{}}
\caption{\label{tab:Timedep-cSTM-components-table} Core components of time-dependent cSTMs with their R name.}\tabularnewline
\toprule
Element & Description & R name & \\
\midrule
\endfirsthead
\toprule
Element & Description & R name & \\
\midrule
\endhead
\(n_S\) & Number of states & \texttt{n\_states} & \\
\(\mathbf{m}_0\) & Initial state vector (row vector) & \texttt{v\_m\_init} & \\
\(\mathbf{m}_t\) & State vector in cycle \(t\) (row vector) & \texttt{v\_mt} & \\
\(M\) & Cohort trace matrix & \texttt{m\_M} & \\
\(\mathbf{P}\) & Time-dependent transition probability array & \texttt{a\_P} & \\
\(\mathbf{A}\) & Transition-dynamics array & \texttt{a\_A} & \\
\(n_{\text{tunnels}}\) & Number of tunnel states & \texttt{n\_tunnel\_size} & \\
\(n_{S_{\text{tunnels}}}\) & Number of states including tunnel states & \texttt{n\_states\_tunnels} & \\
\(\mathbf{m}_{\text{tunnels}_0}\) & Initial state vector for the model with tunnel states & \texttt{v\_m\_init\_tunnels} & \\
\bottomrule
\end{longtable}

\hypertarget{case-study-a-cost-effectiveness-analysis-using-a-time-dependent-sick-sicker-model}{%
\section{Case study: a cost-effectiveness analysis using a time-dependent Sick-Sicker model}\label{case-study-a-cost-effectiveness-analysis-using-a-time-dependent-sick-sicker-model}}

We demonstrate simulation-time and state-residence time-dependence in R by expanding the time-independent 4-state ``Sick-Sicker'' model\textsuperscript{\protect\hyperlink{ref-Enns2015e}{12}} described in the introductory tutorial.\textsuperscript{\protect\hyperlink{ref-Alarid-Escudero2022b}{7}}  We first modify this cSTM to account for simulation-time dependence by incorporating age-dependent background mortality. We then expand it to account for state-residence time dependence in the ``Sick'' state. We will use the time-dependent cSTM to conduct a CEA of different treatment strategies, which we will also use to demonstrate the implementation of transition rewards in cSTM-based CEAs.  We assume the cycle length is one year for the case study, but modelers can adjust the transition probabilities and rewards for a different cycle length. A detailed description of determining the cycle length has been described in the literature.\textsuperscript{\protect\hyperlink{ref-OMahony2015}{13}}

In the introductory tutorial, healthy individuals faced a constant risk of death. In this tutorial, we assume that healthy individuals face an \emph{age-specific} background mortality. To account for the unique quality-of-life impacts and health care needs at the initial onset of the illness, we include a one-time utility decrement of 0.01 (\texttt{du\_HS1}) and a transition cost of \$1,000 (\texttt{ic\_HS1}) for each person making the transition from H to S1 in a given cycle. We assume that when individuals die (i.e., transition to ``D'' ), they incur a one-time cost of \$2,000 (\texttt{ic\_D}) for the expected acute care received right before dying. These impacts are included to illustrate the implementation of transition rewards.

We use this cSTM to evaluate the cost-effectiveness of different treatment strategies involving up to two different treatments (Treatment A and Treatment B) for the Sick-Sicker disease. We assume that it is not clinically possible to distinguish between individuals in S1 and S2, so any treatment strategy involves treating individuals in both S1 and S2, regardless of whether they experience a benefit. Treatment A increases the QoL of individuals only in S1 from 0.75 (utility without treatment, \texttt{u\_S1}) to 0.95 (utility with Treatment A, \texttt{u\_trtA}), costs \$12,000 per year (\texttt{c\_trtA}), and does not impact the transition rates. Treatment B reduces the progression rate from S1 to S2, with a hazard ratio (HR) of 0.6 (\texttt{hr\_S1S2\_trtB}), costs \$13,000 per year (\texttt{c\_trtB}), and does not affect QoL. We consider four treatment strategies: Standard of care with no treatment (SoC Strategy), Treatment A only (Strategy A), Treatment B only (Strategy B), and Treatment A and B used in combination (Strategy AB). When combined, the effects of Treatment A and B are assumed to be independent and treatment costs are additive. Note that for Strategy A, the model has identical transition probabilities to SoC. The differences are the added cost of the treatment for S1 and S2, and QoL increases for S1. After comparing the four strategies in terms of expected QALYs and costs, we calculate the incremental cost per QALY gained between non-dominated strategies as explained below.

Model parameters and the corresponding R variable names are presented in Table \ref{tab:param-table}. We follow the notation described in the DARTH coding framework.\textsuperscript{\protect\hyperlink{ref-Alarid-Escudero2019e}{14}}

\begin{longtable}[]{@{}
  >{\raggedright\arraybackslash}p{(\columnwidth - 6\tabcolsep) * \real{0.45}}
  >{\centering\arraybackslash}p{(\columnwidth - 6\tabcolsep) * \real{0.16}}
  >{\centering\arraybackslash}p{(\columnwidth - 6\tabcolsep) * \real{0.19}}
  >{\centering\arraybackslash}p{(\columnwidth - 6\tabcolsep) * \real{0.20}}@{}}
\caption{\label{tab:param-table} Description of parameters, their R variable name, base-case value and distribution.}\tabularnewline
\toprule
\textbf{Parameter} & \textbf{R name} & \textbf{Base-case} & \textbf{Distribution} \\
\midrule
\endfirsthead
\toprule
\textbf{Parameter} & \textbf{R name} & \textbf{Base-case} & \textbf{Distribution} \\
\midrule
\endhead
Number of cycles (\(n_{T}\)) & \texttt{n\_cycles} & 75 years & - \\
Names of health states & \texttt{v\_names\_states} & H, S1, S2, D & - \\
Annual discount rate for costs & \texttt{d\_c} & 3\% & - \\
Annual discount rate for QALYs & \texttt{d\_e} & 3\% & - \\
Number of PSA samples (\(K\)) & \texttt{n\_sim} & 1,000 & - \\
Annual transition rates & & & \\
- Disease onset (H to S1) & \texttt{r\_HS1} & 0.15 & gamma(30, 200) \\
- Recovery (S1 to H) & \texttt{r\_S1H} & 0.5 & gamma(60, 120) \\
- Time-independent disease progression (S1 to S2) & \texttt{r\_S1S2} & 0.105 & gamma(84, 800) \\
- Time-dependent disease progression (S1 to S2) & \texttt{v\_r\_S1S2\_tunnels} & & \\
~~~~Weibull parameters & & & \\
~~~~~~~~Scale (\(\lambda\)) & \texttt{r\_S1S2\_scale} & 0.08 & MVLN(\(\mu=(0.08, 1.10)\), \\
~~~~~~~~Shape (\(\gamma\)) & \texttt{r\_S1S2\_shape} & 1.10 & \(\sigma=(0.02, 0.05)\), \(\rho=0.5\))$^*$ \\
Annual mortality & & & \\
- Age-dependent background mortality rate (H to D) & \texttt{v\_r\_HDage} & age-specific & - \\
- Hazard ratio of death in S1 vs H & \texttt{hr\_S1} & 3.0 & lognormal(log(3.0), 0.01) \\
- Hazard ratio of death in S2 vs H & \texttt{hr\_S2} & 10.0 & lognormal(log(10.0), 0.02) \\
Annual costs & & & \\
- Healthy individuals & \texttt{c\_H} & \$2,000 & gamma(100.0, 20.0) \\
- Sick individuals in S1 & \texttt{c\_S1} & \$4,000 & gamma(177.8, 22.5) \\
- Sick individuals in S2 & \texttt{c\_S2} & \$15,000 & gamma(225.0, 66.7) \\
- Dead individuals & \texttt{c\_D} & \$0 & - \\
Utility weights & & & \\
- Healthy individuals & \texttt{u\_H} & 1.00 & beta(200, 3) \\
- Sick individuals in S1 & \texttt{u\_S1} & 0.75 & beta(130, 45) \\
- Sick individuals in S2 & \texttt{u\_S2} & 0.50 & beta(230, 230) \\
- Dead individuals & \texttt{u\_D} & 0.00 & - \\
Treatment A: cost and effectiveness & & & \\
- Cost of Treatment A, additional to state-specific health care costs & \texttt{c\_trtA} & \$12,000 & gamma(73.5, 163.3) \\
- Utility for treated individuals in S1 & \texttt{u\_trtA} & 0.95 & beta(300, 15) \\
Treatment B: cost and effectiveness & & & \\
- Cost of Treatment B, additional to state-specific health care costs & \texttt{c\_trtB} & \$12,000 & gamma(86.2, 150.8) \\
- Reduction in rate of disease progression (S1 to S2) as hazard ratio (HR) & \texttt{hr\_S1S2\_trtB} & log(0.6) & lognormal(log(0.6), 0.02) \\
Transition rewards & & & \\
- Utility decrement of healthy individuals & \texttt{du\_HS1} & 0.01 & beta(11,1088) \\
when transitioning to S1 & & & \\
- Cost of healthy individuals & \texttt{ic\_HS1} & \$1,000 & gamma(25, 40) \\
when transitioning to S1 & & & \\
- Cost of dying when transitioning to D & \texttt{ic\_D} & \$2,000 & gamma(100, 20) \\
\bottomrule
$^*$MVLN: Multivariate log-normal distribution
\end{longtable}

\hypertarget{incorporating-simulation-time-dependence}{%
\subsection{Incorporating simulation-time dependence}\label{incorporating-simulation-time-dependence}}

To illustrate simulation-time dependence in the Sick-Sicker cSTM, we model all-cause mortality as a function of age. We obtain all-cause mortality from life tables in the form of age-specific mortality hazard rates, \(\mu(a)\), where \(a\) refers to age. For this example, we create a vector \texttt{v\_r\_mort\_by\_age} to represent age-specific background mortality hazard rates for 25 to 100 year-olds obtained from US life tables.\textsuperscript{\protect\hyperlink{ref-Arias2017}{15}} The .csv file with the life tables is included in the Supplementary Material and the GitHub repository. We repeat each of the one-year-specific rates in \(\mu(a)\) (R variable name \texttt{v\_r\_mort\_by\_age}) as many times as the number of cycles in a year for these ages. To compute the transition probability from state H to state D, corresponding to the cohort's age at each cycle, we transform the rate \(\mu(a)\) to a transition probability assuming a constant exponential hazard rate within each year of age
\[
  p_{[H,D,t]} = 1-\exp\left\{{-\mu(a_0 + t)}\right\},
\]
where \(a_0 =\) 25 is the starting age of the cohort. We transform the resulting R variable, \texttt{v\_r\_HDage}, to a vector of probabilities, \texttt{v\_p\_HDage}, adjusting by the cycle length.

\begin{Shaded}
\begin{Highlighting}[]
\DocumentationTok{\#\# Load life table mortality rates from \textasciigrave{}.csv\textasciigrave{} format}
\NormalTok{lt\_usa\_2005 }\OtherTok{\textless{}{-}} \FunctionTok{read.csv}\NormalTok{(}\StringTok{"LifeTable\_USA\_Mx\_2015.csv"}\NormalTok{)}
\CommentTok{\# Extract age{-}specific all{-}cause mortality for ages in model time horizon}
\NormalTok{v\_r\_mort\_by\_age }\OtherTok{\textless{}{-}}\NormalTok{ lt\_usa\_2015 }\SpecialCharTok{\%\textgreater{}\%} 
\NormalTok{  dplyr}\SpecialCharTok{::}\FunctionTok{filter}\NormalTok{(Age }\SpecialCharTok{\textgreater{}=}\NormalTok{ n\_age\_init }\SpecialCharTok{\&}\NormalTok{ Age }\SpecialCharTok{\textless{}}\NormalTok{ n\_age\_max) }\SpecialCharTok{\%\textgreater{}\%}
\NormalTok{  dplyr}\SpecialCharTok{::}\FunctionTok{select}\NormalTok{(Total) }\SpecialCharTok{\%\textgreater{}\%}
  \FunctionTok{as.matrix}\NormalTok{()}
\CommentTok{\# Age{-}specific mortality rate in the Healthy state (background mortality) }
\CommentTok{\# for all cycles}
\NormalTok{v\_r\_HDage  }\OtherTok{\textless{}{-}} \FunctionTok{rep}\NormalTok{(v\_r\_mort\_by\_age, }\AttributeTok{each =} \DecValTok{1}\SpecialCharTok{/}\NormalTok{cycle\_length)}
\CommentTok{\# Transform to age{-}specific background mortality risk for all cycles adjusting }
\CommentTok{\# by cycle length}
\NormalTok{v\_p\_HDage  }\OtherTok{\textless{}{-}} \DecValTok{1} \SpecialCharTok{{-}} \FunctionTok{exp}\NormalTok{(}\SpecialCharTok{{-}}\NormalTok{v\_r\_HDage }\SpecialCharTok{*}\NormalTok{ cycle\_length) }
\end{Highlighting}
\end{Shaded}

Because mortality in S1 and S2 are relative to background mortality which depends on age, mortality in S1 and S2 will also be age-dependent. To generate the age-specific mortality in S1 and S2, we multiply the age-specific background mortality rate, \texttt{v\_r\_HDage}, by the constant hazard ratios \texttt{hr\_S1} and \texttt{hr\_S2}, respectively. We then convert the resulting age-specific mortality rates to probabilities.

\begin{Shaded}
\begin{Highlighting}[]
\DocumentationTok{\#\# Age{-}specific mortality rates in the Sick and Sicker states}
\NormalTok{v\_r\_S1Dage }\OtherTok{\textless{}{-}}\NormalTok{ v\_r\_HDage }\SpecialCharTok{*}\NormalTok{ hr\_S1 }\CommentTok{\# when Sick}
\NormalTok{v\_r\_S2Dage }\OtherTok{\textless{}{-}}\NormalTok{ v\_r\_HDage }\SpecialCharTok{*}\NormalTok{ hr\_S2 }\CommentTok{\# when Sicker}
\DocumentationTok{\#\# Age{-}specific probabilities of dying in the Sick and Sicker states}
\NormalTok{v\_p\_S1Dage }\OtherTok{\textless{}{-}} \DecValTok{1} \SpecialCharTok{{-}} \FunctionTok{exp}\NormalTok{(}\SpecialCharTok{{-}}\NormalTok{v\_r\_S1Dage) }\CommentTok{\# when Sick}
\NormalTok{v\_p\_S2Dage }\OtherTok{\textless{}{-}} \DecValTok{1} \SpecialCharTok{{-}} \FunctionTok{exp}\NormalTok{(}\SpecialCharTok{{-}}\NormalTok{v\_r\_S2Dage) }\CommentTok{\# when Sicker}
\end{Highlighting}
\end{Shaded}

\begin{figure}[H]

{\centering \includegraphics[width=1\linewidth]{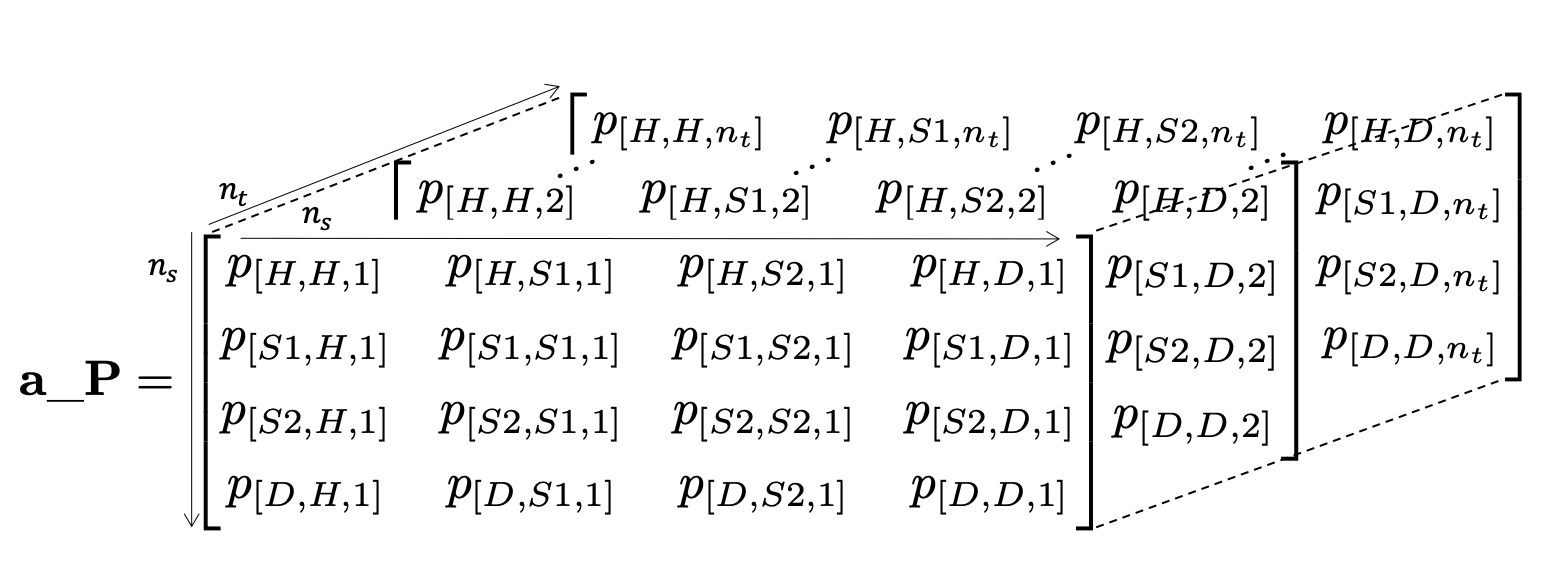} 

}

\caption{A 3-dimensional representation of the transition probability array of the Sick-Sicker model with simulation-time dependence.}\label{fig:Array-Time-Dependent}
\end{figure}

To incorporate simulation-time dependence into the transition probability matrix, we expand the dimensions of the matrix and create a 3-dimensional transition probability array, \(\mathbf{P}\) and use \texttt{a\_P} in R, of dimensions \(n_S \times n_S \times n_T\). The first two dimensions of this array correspond to transitions between states, where the rows determine departing states and columns destination states. The third dimension refers to the time since the simulation started. The third dimension refers to time since simulation start. The \(t\)-th element in the third dimension corresponds to the transition probability matrix at cycle \(t\). A visual representation of \texttt{a\_P} is shown in Figure \ref{fig:Array-Time-Dependent}.

First, we initialize the transition probability array for SoC, \texttt{a\_P\_SoC}, with a default value of zero for all transition probabilities.

\begin{Shaded}
\begin{Highlighting}[]
\CommentTok{\# Initialize the transition probability array}
\NormalTok{a\_P\_SoC }\OtherTok{\textless{}{-}} \FunctionTok{array}\NormalTok{(}\DecValTok{0}\NormalTok{, }\AttributeTok{dim =} \FunctionTok{c}\NormalTok{(n\_states, n\_states, n\_cycles),}
              \AttributeTok{dimnames =} \FunctionTok{list}\NormalTok{(v\_names\_states, v\_names\_states, }\DecValTok{0}\SpecialCharTok{:}\NormalTok{(n\_cycles }\SpecialCharTok{{-}} \DecValTok{1}\NormalTok{)))}
\end{Highlighting}
\end{Shaded}

The code below illustrates how to assign age-dependent transition probabilities in the third dimension of the array \texttt{a\_P\_SoC}.

\begin{Shaded}
\begin{Highlighting}[]
\DocumentationTok{\#\#\# Fill in array}
\DocumentationTok{\#\# From H}
\NormalTok{a\_P\_SoC[}\StringTok{"H"}\NormalTok{, }\StringTok{"H"}\NormalTok{, ]   }\OtherTok{\textless{}{-}}\NormalTok{ (}\DecValTok{1} \SpecialCharTok{{-}}\NormalTok{ v\_p\_HDage) }\SpecialCharTok{*}\NormalTok{ (}\DecValTok{1} \SpecialCharTok{{-}}\NormalTok{ p\_HS1)}
\NormalTok{a\_P\_SoC[}\StringTok{"H"}\NormalTok{, }\StringTok{"S1"}\NormalTok{, ]  }\OtherTok{\textless{}{-}}\NormalTok{ (}\DecValTok{1} \SpecialCharTok{{-}}\NormalTok{ v\_p\_HDage) }\SpecialCharTok{*}\NormalTok{ p\_HS1}
\NormalTok{a\_P\_SoC[}\StringTok{"H"}\NormalTok{, }\StringTok{"D"}\NormalTok{, ]   }\OtherTok{\textless{}{-}}\NormalTok{ v\_p\_HDage}
\DocumentationTok{\#\# From S1}
\NormalTok{a\_P\_SoC[}\StringTok{"S1"}\NormalTok{, }\StringTok{"H"}\NormalTok{, ]  }\OtherTok{\textless{}{-}}\NormalTok{ (}\DecValTok{1} \SpecialCharTok{{-}}\NormalTok{ v\_p\_S1Dage) }\SpecialCharTok{*}\NormalTok{ p\_S1H}
\NormalTok{a\_P\_SoC[}\StringTok{"S1"}\NormalTok{, }\StringTok{"S1"}\NormalTok{, ] }\OtherTok{\textless{}{-}}\NormalTok{ (}\DecValTok{1} \SpecialCharTok{{-}}\NormalTok{ v\_p\_S1Dage) }\SpecialCharTok{*}\NormalTok{ (}\DecValTok{1} \SpecialCharTok{{-}}\NormalTok{ (p\_S1H }\SpecialCharTok{+}\NormalTok{ p\_S1S2))}
\NormalTok{a\_P\_SoC[}\StringTok{"S1"}\NormalTok{, }\StringTok{"S2"}\NormalTok{, ] }\OtherTok{\textless{}{-}}\NormalTok{ (}\DecValTok{1} \SpecialCharTok{{-}}\NormalTok{ v\_p\_S1Dage) }\SpecialCharTok{*}\NormalTok{ p\_S1S2}
\NormalTok{a\_P\_SoC[}\StringTok{"S1"}\NormalTok{, }\StringTok{"D"}\NormalTok{, ]  }\OtherTok{\textless{}{-}}\NormalTok{ v\_p\_S1Dage}
\DocumentationTok{\#\# From S2}
\NormalTok{a\_P\_SoC[}\StringTok{"S2"}\NormalTok{, }\StringTok{"S2"}\NormalTok{, ] }\OtherTok{\textless{}{-}} \DecValTok{1} \SpecialCharTok{{-}}\NormalTok{ v\_p\_S2Dage}
\NormalTok{a\_P\_SoC[}\StringTok{"S2"}\NormalTok{, }\StringTok{"D"}\NormalTok{, ]  }\OtherTok{\textless{}{-}}\NormalTok{ v\_p\_S2Dage}
\DocumentationTok{\#\# From D}
\NormalTok{a\_P\_SoC[}\StringTok{"D"}\NormalTok{, }\StringTok{"D"}\NormalTok{, ]   }\OtherTok{\textless{}{-}} \DecValTok{1}

\DocumentationTok{\#\# Initialize transition probability matrix for Strategy A as a copy of SoC\textquotesingle{}s}
\NormalTok{a\_P\_strA }\OtherTok{\textless{}{-}}\NormalTok{ a\_P\_SoC}
\end{Highlighting}
\end{Shaded}

Most of this code is equivalent to populating the matrix \(P\) for time-independent cSTM. The only difference is the empty index added to the third dimension of \texttt{a\_P\_SoC} to represent the vector of transitions over time. For time-varying transitions, we provide a vector of transition probabilities. We only need to provide one value for the transition probability for constant transitions over time, and R replicates the value of such transitions as many times as the number of cycles (\(n_T+1\) times in our example). To ensure that the sum of each row of each cycle-specific matrix equals one, we define the probability of staying in each state by subtracting the sum of the exiting probability vectors from one after conditioning on surviving each cycle. Finally, we initialize the transition probability array for Strategy A as a copy of SoC's because Treatment A does not alter the cohort's transition probabilities.

Each slice along the third dimension of \texttt{a\_P\_SoC} corresponds to a transition probability matrix from \(t\) to \(t+1\). For example, the transition matrix for 25-year-olds in the Sick-Sicker model under the SoC Strategy can be retrieved by indexing the first slice of the array using:

\begin{Shaded}
\begin{Highlighting}[]
\NormalTok{a\_P\_SoC[, , }\DecValTok{1}\NormalTok{]}
\end{Highlighting}
\end{Shaded}

\begin{verbatim}
##            H        S1        S2           D
## H  0.8491385 0.1498480 0.0000000 0.001013486
## S1 0.4984813 0.3938002 0.1046811 0.003037378
## S2 0.0000000 0.0000000 0.9899112 0.010088764
## D  0.0000000 0.0000000 0.0000000 1.000000000
\end{verbatim}

Similarly, the code below defines \texttt{a\_P\_strB} for Strategy B.

\begin{Shaded}
\begin{Highlighting}[]
\DocumentationTok{\#\# Initialize transition probability array for Strategy B}
\NormalTok{a\_P\_strB }\OtherTok{\textless{}{-}}\NormalTok{ a\_P\_SoC}
\DocumentationTok{\#\# Update only transition probabilities from S1 involving p\_S1S2}
\NormalTok{a\_P\_strB[}\StringTok{"S1"}\NormalTok{, }\StringTok{"S1"}\NormalTok{, ] }\OtherTok{\textless{}{-}}\NormalTok{ (}\DecValTok{1} \SpecialCharTok{{-}}\NormalTok{ v\_p\_S1Dage) }\SpecialCharTok{*}\NormalTok{ (}\DecValTok{1} \SpecialCharTok{{-}}\NormalTok{ (p\_S1H }\SpecialCharTok{+}\NormalTok{ p\_S1S2\_trtB))}
\NormalTok{a\_P\_strB[}\StringTok{"S1"}\NormalTok{, }\StringTok{"S2"}\NormalTok{, ] }\OtherTok{\textless{}{-}}\NormalTok{ (}\DecValTok{1} \SpecialCharTok{{-}}\NormalTok{ v\_p\_S1Dage) }\SpecialCharTok{*}\NormalTok{ p\_S1S2\_trtB}

\DocumentationTok{\#\# Initialize transition probability matrix for Strategy AB as a copy of B\textquotesingle{}s}
\NormalTok{a\_P\_strAB }\OtherTok{\textless{}{-}}\NormalTok{ a\_P\_strB}
\end{Highlighting}
\end{Shaded}

We instantiate \texttt{a\_P\_strB} as a copy of \texttt{a\_P\_SoC} and update the probability of remaining in S1, and the transition probability from S1 to S2 (i.e., \texttt{p\_S1S2} is replaced with \texttt{p\_S1S2\_trtB}). Next, we create the transition probability array for Strategy AB, \texttt{a\_P\_strAB}, as a copy of \texttt{a\_P\_strB} since the cSTMs for Strategies B and AB have identical transition probabilities but are only different in cost, as illustrated below.

We used the functions \texttt{check\_sum\_of\_transition\_array} and \texttt{check\_transition\_probability} in the \texttt{darthtools} package (\url{https://github.com/DARTH-git/darthtools}) to check if the transition probability arrays are valid (i.e., ensuring transition probabilities are between 0 and 1, and transition probabilities from each state sum to 1). The Supplementary Material includes the code for these functions.

In the Sick-Sicker model, the entire cohort starts in H.The creation of the initial state vector, \texttt{v\_m\_init}, and the cohort trace, \texttt{m\_M}, for all strategies is identical to the time-independent cSTM in the introductory tutorial.\textsuperscript{\protect\hyperlink{ref-Alarid-Escudero2022b}{7}} The code below iteratively computes the cohort's distribution across health states over time for the SoC Strategy.

\begin{Shaded}
\begin{Highlighting}[]
\CommentTok{\# Iterative solution of age{-}dependent cSTM under SoC}
\ControlFlowTok{for}\NormalTok{(t }\ControlFlowTok{in} \DecValTok{1}\SpecialCharTok{:}\NormalTok{n\_cycles)\{}
\NormalTok{  m\_M\_SoC[t }\SpecialCharTok{+} \DecValTok{1}\NormalTok{, ] }\OtherTok{\textless{}{-}}\NormalTok{ m\_M\_SoC[t, ] }\SpecialCharTok{\%*\%}\NormalTok{ a\_P\_SoC[, , t]}
\NormalTok{\}}
\end{Highlighting}
\end{Shaded}

\begin{figure}[H]

{\centering \includegraphics{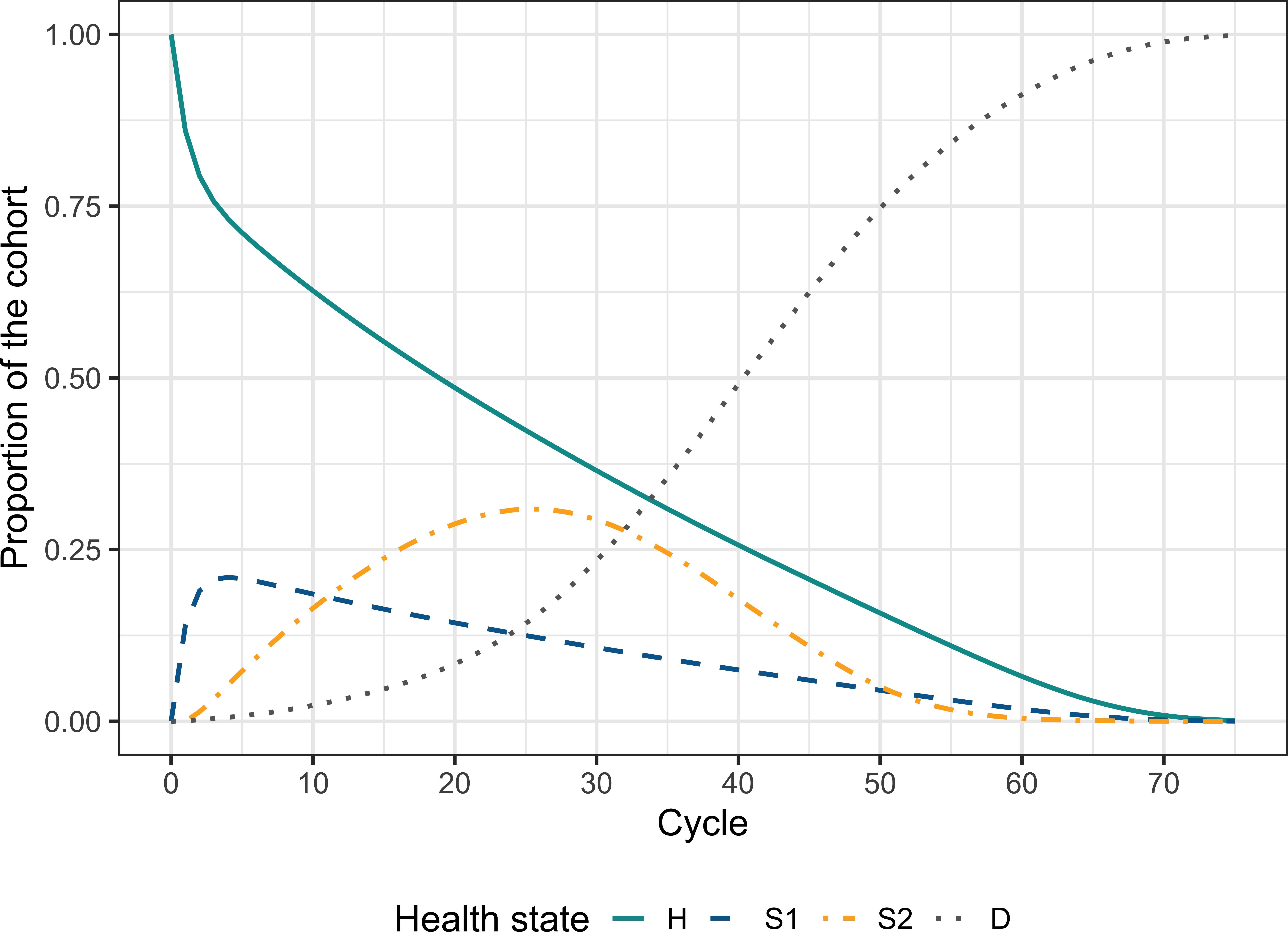} 

}

\caption{Cohort trace of the age-dependent cSTM under SoC.}\label{fig:Sick-Sicker-Trace-AgeDep}
\end{figure}

The state vector of the cohort's distribution at cycle \(t+1\) is obtained by applying Equation \eqref{eq:time-dep-matrix-mult}, the matrix product between the row vector \texttt{m\_M\_SoC{[}t,\ {]}} and the transition probability matrix \texttt{a\_P\_SoC{[},\ ,\ t{]}}. This equation is comparable to the one described for the time-independent model in the introductory tutorial.\textsuperscript{\protect\hyperlink{ref-Alarid-Escudero2022b}{7}} The only modification required is to index the third dimension of the transition probability arrays by \(t\) to obtain the SoC's cycle-specific transition probability matrix. The vectors across all cycles are obtained by iteratively applying Equation \eqref{eq:time-dep-matrix-mult} to each transition probability matrix of \texttt{a\_P\_SoC} across the third dimension using a \texttt{for} loop. Figure \ref{fig:Sick-Sicker-Trace-AgeDep} shows the cohort's trace for all cycles of the age-dependent cSTM under SoC. The complete code for all the strategies is included in the Supplementary Material.

\hypertarget{incorporating-time-dependence-on-state-residence}{%
\subsection{Incorporating time dependence on state residence (tunnel states)}\label{incorporating-time-dependency-on-state-residence}}

We add the dependence on state-residence to the simulation-time-dependent Sick-Sicker model defined above. We assume the risk of progression from S1 to S2 increases as a function of the time \(\tau = 1, \ldots, n_{\text{tunnels}}\) the cohort remains in the S1 state. This increase follows a Weibull hazard function, \(h(\tau)\), defined as
\[
  h(\tau) = \gamma \lambda (\lambda \tau)^{\gamma-1},
\]
with a corresponding cumulative hazard, \(H(\tau)\),
\begin{equation}
  H(\tau) = (\lambda \tau)^{\gamma},
\label{eq:H-weibull}
\end{equation}
where \(\lambda\) and \(\gamma\) are the scale and shape parameters of the Weibull hazard function, respectively.

To derive a transition probability from S1 to S2 as a function of the time the cohort spends in S1, \(p_{\left[S1_{\tau},S2, \tau\right]}\), we assume constant rates within each cycle interval (i.e., piecewise exponential transition times), where the cycle-specific probability of a transition is
\begin{equation}
  p_{\left[S1_{\tau},S2, \tau\right]} = 1-\exp{\left(-\mu_{\left[S1_{\tau},S2, \tau\right]}\right)},
\label{eq:tp-from-rate}
\end{equation}
where \(\mu_{\left[S1_{\tau},S2, \tau\right]}\) is the rate of transition from S1 to S2 in cycle \(\tau\) defined as the difference in cumulative hazards between consecutive cycles\textsuperscript{\protect\hyperlink{ref-Diaby2014}{16}}
\begin{equation}
  \mu_{\left[S1_{\tau},S2, \tau\right]} = H(\tau) - H(\tau-1).
\label{eq:tr-from-H}
\end{equation}

Substituting the Weibull cumulative hazard from Equation \eqref{eq:H-weibull} into Equation \eqref{eq:tr-from-H} gives
\begin{equation}
  \mu_{\left[S1_{\tau},S2, \tau\right]} = (\lambda \tau)^{\gamma} - (\lambda (\tau-1))^{\gamma},
\label{eq:tr-from-H-weibull}
\end{equation}
and the transition probability
\begin{equation}
  p_{\left[S1_{\tau},S2, \tau\right]} = 1-\exp{\left(- \left((\lambda \tau)^{\gamma} - (\lambda (\tau-1))^{\gamma}\right) \right)}.
\label{eq:tp-from-H-weibull}
\end{equation}

We assume that state-residence dependence affects the cohort in the S1 state throughout the whole simulation (i.e., \(n_{\text{tunnels}}=n_T\)) and create a new variable called \texttt{n\_tunnel\_size} with the length of the tunnel equal to \texttt{n\_cycles}. Thus, there will be 75 S1 tunnel states plus 3 more states (H, S2, D) resulting in a total of \(n_{S_{\text{tunnels}}}\) = 78.

\begin{figure}[H]

{\centering \includegraphics[width=1\textwidth]{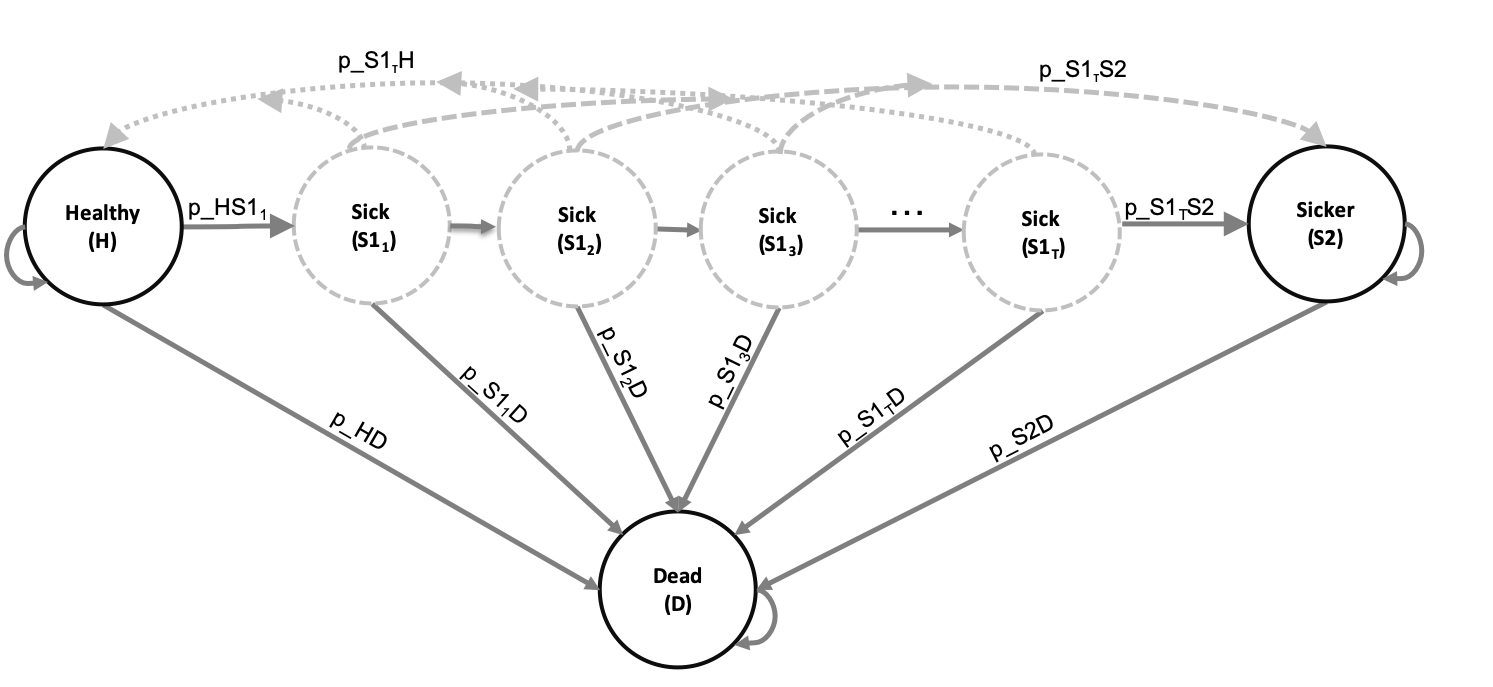} 

}

\caption{State-transition diagram of the Sick-Sicker model with tunnel states expanding the Sick state ($S1_1, S1_2,...,S1_{n_\text{tunnels}}$). The circles represent the health states, and the arrows represent the possible transition probabilities. The labels next to the arrows represent the variable names for these transitions.}\label{fig:STD-Sick-Sicker-tunnels}
\end{figure}

Figure \ref{fig:STD-Sick-Sicker-tunnels} shows the state-transition diagram of the Sick-Sicker model with state-residence dependency with \(n_{\text{tunnels}}\) tunnel states for S1.

To implement state-residence dependence in the Sick-Sicker cSTM, we create the vector variables \texttt{v\_Sick\_tunnel} and \texttt{v\_names\_states\_tunnels} with the names of the Sick tunnel states' and all the states of the cSTM, including tunnels, respectively, and use the parameters listed in Tables \ref{tab:Timedep-cSTM-components-table} and \ref{tab:param-table}.

\begin{Shaded}
\begin{Highlighting}[]
\DocumentationTok{\#\# Number of tunnels}
\NormalTok{n\_tunnel\_size }\OtherTok{\textless{}{-}}\NormalTok{ n\_cycles }
\DocumentationTok{\#\# Vector with cycles for tunnels}
\NormalTok{v\_cycles\_tunnel }\OtherTok{\textless{}{-}} \DecValTok{1}\SpecialCharTok{:}\NormalTok{n\_tunnel\_size}
\DocumentationTok{\#\# Vector with the names of the Sick tunnel state}
\NormalTok{v\_Sick\_tunnel }\OtherTok{\textless{}{-}} \FunctionTok{paste}\NormalTok{(}\StringTok{"S1\_"}\NormalTok{, }\FunctionTok{seq}\NormalTok{(}\DecValTok{1}\NormalTok{, n\_tunnel\_size), }\StringTok{"Yr"}\NormalTok{, }\AttributeTok{sep =} \StringTok{""}\NormalTok{)}
\DocumentationTok{\#\# Create variables for model with tunnels}
\NormalTok{v\_names\_states\_tunnels }\OtherTok{\textless{}{-}} \FunctionTok{c}\NormalTok{(}\StringTok{"H"}\NormalTok{, v\_Sick\_tunnel, }\StringTok{"S2"}\NormalTok{, }\StringTok{"D"}\NormalTok{) }\CommentTok{\# state names}
\NormalTok{n\_states\_tunnels }\OtherTok{\textless{}{-}} \FunctionTok{length}\NormalTok{(v\_names\_states\_tunnels)         }\CommentTok{\# number of states}
\DocumentationTok{\#\# Initialize first cycle of Markov trace accounting for the tunnels}
\NormalTok{v\_m\_init\_tunnels }\OtherTok{\textless{}{-}} \FunctionTok{c}\NormalTok{(}\DecValTok{1}\NormalTok{, }\FunctionTok{rep}\NormalTok{(}\DecValTok{0}\NormalTok{, n\_tunnel\_size), }\DecValTok{0}\NormalTok{, }\DecValTok{0}\NormalTok{) }
\end{Highlighting}
\end{Shaded}

Then, the transition rate and probability dependent on state residence from Sick to Sicker, \texttt{v\_r\_S1S2\_tunnels} and \texttt{v\_p\_S1S2\_tunnels}, respectively, based on a Weibull hazard function are:

\begin{Shaded}
\begin{Highlighting}[]
\CommentTok{\# Weibull parameters}
\NormalTok{r\_S1S2\_scale }\OtherTok{\textless{}{-}} \FloatTok{0.08} \CommentTok{\# scale}
\NormalTok{r\_S1S2\_shape }\OtherTok{\textless{}{-}} \FloatTok{1.10} \CommentTok{\# shape}
\CommentTok{\# Weibull function}
\NormalTok{v\_r\_S1S2\_tunnels }\OtherTok{\textless{}{-}}\NormalTok{ (v\_cycles\_tunnel}\SpecialCharTok{*}\NormalTok{r\_S1S2\_scale)}\SpecialCharTok{\^{}}\NormalTok{r\_S1S2\_shape }\SpecialCharTok {{-}}
\NormalTok{                    ((v\_cycles\_tunnel} \DecValTok{{-} 1}\NormalTok{)}\SpecialCharTok{*}\NormalTok{r\_S1S2\_scale)}\SpecialCharTok{\^{}}\NormalTok{r\_S1S2\_shape}
                    
\NormalTok{v\_p\_S1S2\_tunnels }\OtherTok{\textless{}{-}} \DecValTok{1} \SpecialCharTok{{-}} \FunctionTok{exp}\NormalTok{(}\SpecialCharTok{{-}}\NormalTok{v\_r\_S1S2\_tunnels}\SpecialCharTok{*}\NormalTok{cycle\_length)}
\end{Highlighting}
\end{Shaded}

\begin{figure}[H]

{\centering \includegraphics[width=1\textwidth]{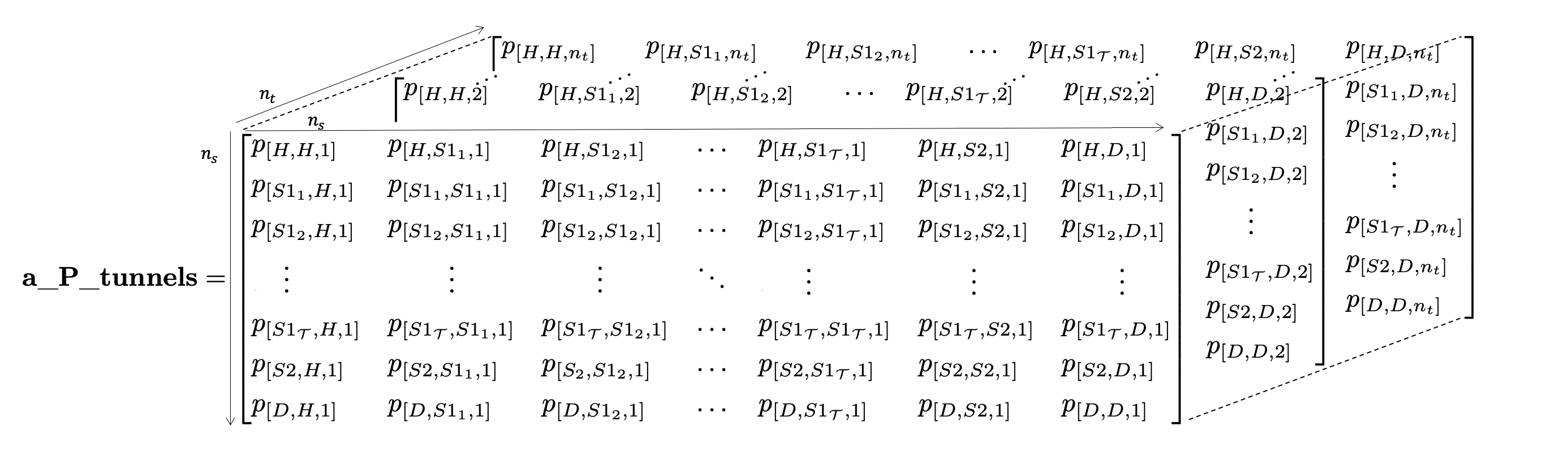} 

}

\caption{The 3-dimensional transition probability array of the Sick-Sicker model expanded to account for simulation-time and state-residence time dependence using $\tau$ tunnel states for S1.}\label{fig:Array-Time-Dependent-Tunnels}
\end{figure}

To adapt the 3-dimensional transition probability array to incorporate both age and state-residence dependence in the Sick-Sicker model under SoC, we first create an expanded 3-dimensional array accounting for tunnels, \texttt{a\_P\_tunnels\_SoC}. The dimensions of this array are \(n_{S_{\text{tunnels}}} \times n_{S_{\text{tunnels}}} \times n_T\). A visual representation of \texttt{a\_P\_tunnels\_SoC} of the Sick-Sicker model with tunnel states expanding the Sick state is shown in Figure \ref{fig:Array-Time-Dependent-Tunnels}.

\begin{Shaded}
\begin{Highlighting}[]
\CommentTok{\# Initialize array}
\NormalTok{a\_P\_tunnels\_SoC }\OtherTok{\textless{}{-}} \FunctionTok{array}\NormalTok{(}\DecValTok{0}\NormalTok{, }\AttributeTok{dim =} \FunctionTok{c}\NormalTok{(n\_states\_tunnels, n\_states\_tunnels, n\_cycles),}
                         \AttributeTok{dimnames =} \FunctionTok{list}\NormalTok{(v\_names\_states\_tunnels, }
\NormalTok{                                         v\_names\_states\_tunnels, }
                                         \DecValTok{0}\SpecialCharTok{:}\NormalTok{(n\_cycles }\SpecialCharTok{{-}} \DecValTok{1}\NormalTok{)))}
\end{Highlighting}
\end{Shaded}

Filling \texttt{a\_P\_tunnels\_SoC} with the corresponding transition probabilities works similarly to \texttt{a\_P\_SoC} above. The difference is that we now fill the transition probabilities to/from tunnel states by iterating over the tunnel and assigning the appropriate disease progression transition probabilities for each state as described in Section \ref{time-dependence-on-state-residence} and Algorithm \ref{alg:fill-tunnels}.

\begin{Shaded}
\begin{Highlighting}[]
\DocumentationTok{\#\#\# Fill in array}
\DocumentationTok{\#\# From H}
\NormalTok{a\_P\_tunnels\_SoC[}\StringTok{"H"}\NormalTok{, }\StringTok{"H"}\NormalTok{, ]              }\OtherTok{\textless{}{-}}\NormalTok{ (}\DecValTok{1} \SpecialCharTok{{-}}\NormalTok{ v\_p\_HDage) }\SpecialCharTok{*}\NormalTok{ (}\DecValTok{1} \SpecialCharTok{{-}}\NormalTok{ p\_HS1)}
\NormalTok{a\_P\_tunnels\_SoC[}\StringTok{"H"}\NormalTok{, v\_Sick\_tunnel[}\DecValTok{1}\NormalTok{], ] }\OtherTok{\textless{}{-}}\NormalTok{ (}\DecValTok{1} \SpecialCharTok{{-}}\NormalTok{ v\_p\_HDage) }\SpecialCharTok{*}\NormalTok{ p\_HS1}
\NormalTok{a\_P\_tunnels\_SoC[}\StringTok{"H"}\NormalTok{, }\StringTok{"D"}\NormalTok{, ]              }\OtherTok{\textless{}{-}}\NormalTok{ v\_p\_HDage}
\DocumentationTok{\#\# From S1}
\ControlFlowTok{for}\NormalTok{ (i }\ControlFlowTok{in} \DecValTok{1}\SpecialCharTok{:}\NormalTok{n\_tunnel\_size) \{}
\NormalTok{  a\_P\_tunnels\_SoC[v\_Sick\_tunnel[i], }\StringTok{"H"}\NormalTok{, ]  }\OtherTok{\textless{}{-}}\NormalTok{ (}\DecValTok{1} \SpecialCharTok{{-}}\NormalTok{ v\_p\_S1Dage) }\SpecialCharTok{*}\NormalTok{ p\_S1H}
\NormalTok{  a\_P\_tunnels\_SoC[v\_Sick\_tunnel[i], }\StringTok{"S2"}\NormalTok{, ] }\OtherTok{\textless{}{-}}\NormalTok{ (}\DecValTok{1} \SpecialCharTok{{-}}\NormalTok{ v\_p\_S1Dage) }\SpecialCharTok{*}\NormalTok{ v\_p\_S1S2\_tunnels[i]}
\NormalTok{  a\_P\_tunnels\_SoC[v\_Sick\_tunnel[i], }\StringTok{"D"}\NormalTok{, ]  }\OtherTok{\textless{}{-}}\NormalTok{ v\_p\_S1Dage}
  \ControlFlowTok{if}\NormalTok{ (i }\SpecialCharTok{==}\NormalTok{ n\_tunnel\_size) \{}
    \CommentTok{\# If reaching last tunnel state, ensure that the cohort that does not}
    \CommentTok{\# transition out of the Sick state, remain in the last Sick tunnel state}
\NormalTok{    a\_P\_tunnels\_SoC[v\_Sick\_tunnel[i],}
\NormalTok{                    v\_Sick\_tunnel[i], ] }\OtherTok{\textless{}{-}}\NormalTok{ (}\DecValTok{1} \SpecialCharTok{{-}}\NormalTok{ v\_p\_S1Dage) }\SpecialCharTok{*}
\NormalTok{      (}\DecValTok{1} \SpecialCharTok{{-}}\NormalTok{ (p\_S1H }\SpecialCharTok{+}\NormalTok{ v\_p\_S1S2\_tunnels[i]))}
\NormalTok{  \} }\ControlFlowTok{else}\NormalTok{\{}
\NormalTok{    a\_P\_tunnels\_SoC[v\_Sick\_tunnel[i], }
\NormalTok{                    v\_Sick\_tunnel[i }\SpecialCharTok{+} \DecValTok{1}\NormalTok{], ]   }\OtherTok{\textless{}{-}}\NormalTok{ (}\DecValTok{1} \SpecialCharTok{{-}}\NormalTok{ v\_p\_S1Dage) }\SpecialCharTok{*}
\NormalTok{      (}\DecValTok{1} \SpecialCharTok{{-}}\NormalTok{ (p\_S1H }\SpecialCharTok{+}\NormalTok{ v\_p\_S1S2\_tunnels[i]))}
\NormalTok{  \}}
\NormalTok{\}}
\DocumentationTok{\#\#\# From S2}
\NormalTok{a\_P\_tunnels\_SoC[}\StringTok{"S2"}\NormalTok{, }\StringTok{"S2"}\NormalTok{, ] }\OtherTok{\textless{}{-}} \DecValTok{1} \SpecialCharTok{{-}}\NormalTok{ v\_p\_S2Dage}
\NormalTok{a\_P\_tunnels\_SoC[}\StringTok{"S2"}\NormalTok{, }\StringTok{"D"}\NormalTok{, ]  }\OtherTok{\textless{}{-}}\NormalTok{ v\_p\_S2Dage}
\CommentTok{\# From D}
\NormalTok{a\_P\_tunnels\_SoC[}\StringTok{"D"}\NormalTok{, }\StringTok{"D"}\NormalTok{, ] }\OtherTok{\textless{}{-}} \DecValTok{1}
\end{Highlighting}
\end{Shaded}

Again, we check the three-dimensional transition probability array with tunnels is valid using the \texttt{darthtools} package.

To simulate the cohort and store its state occupation over the \(n_T\) cycles for the cSTM accounting for state-residence dependency, we initialize a cohort trace matrix for the SoC Strategy, \texttt{m\_M\_tunnels\_SoC}. The dimensions of the matrix are \((n_T+1) \times n_{S_{\text{tunnels}}}\).

\begin{Shaded}
\begin{Highlighting}[]
\CommentTok{\# Initialize cohort for state{-}residence cSTM under SoC}
\NormalTok{m\_M\_tunnels\_SoC }\OtherTok{\textless{}{-}} \FunctionTok{matrix}\NormalTok{(}\DecValTok{0}\NormalTok{, }
                      \AttributeTok{nrow =}\NormalTok{ (n\_cycles }\SpecialCharTok{+} \DecValTok{1}\NormalTok{), }\AttributeTok{ncol =}\NormalTok{ n\_states\_tunnels, }
                      \AttributeTok{dimnames =} \FunctionTok{list}\NormalTok{(}\DecValTok{0}\SpecialCharTok{:}\NormalTok{n\_cycles, v\_names\_states\_tunnels))}
\CommentTok{\# Store the initial state vector in the first row of the cohort trace}
\NormalTok{m\_M\_tunnels\_SoC[}\DecValTok{1}\NormalTok{, ] }\OtherTok{\textless{}{-}}\NormalTok{ v\_m\_init\_tunnels}
\end{Highlighting}
\end{Shaded}

We then use the matrix product, similar to the simulation-time-dependent cSTM, to generate the full cohort trace.

\begin{Shaded}
\begin{Highlighting}[]
\CommentTok{\# Iterative solution of state{-}residence{-}dependent cSTM under SoC}
\ControlFlowTok{for}\NormalTok{(t }\ControlFlowTok{in} \DecValTok{1}\SpecialCharTok{:}\NormalTok{n\_cycles)\{}
\NormalTok{  m\_M\_tunnels\_SoC[t }\SpecialCharTok{+} \DecValTok{1}\NormalTok{, ] }\OtherTok{\textless{}{-}}\NormalTok{ m\_M\_tunnels\_SoC[t, ] }\SpecialCharTok{\%*\%}\NormalTok{ a\_P\_tunnels\_SoC[, , t]}
\NormalTok{\}}
\end{Highlighting}
\end{Shaded}

To compute a summarized cohort trace to capture occupancy in the H, S1, S2, D states under SoC, we aggregate over the S1 tunnel states in each cycle. The complete code for all the strategies is included in the Supplementary Material.

\begin{Shaded}
\begin{Highlighting}[]
\CommentTok{\# Create aggregated trace}
\NormalTok{m\_M\_tunnels\_SoC\_sum }\OtherTok{\textless{}{-}} \FunctionTok{cbind}\NormalTok{(}\AttributeTok{H =}\NormalTok{ m\_M\_tunnels\_SoC[, }\StringTok{"H"}\NormalTok{], }
                             \AttributeTok{S1 =} \FunctionTok{rowSums}\NormalTok{(m\_M\_tunnels\_SoC[, }\FunctionTok{which}\NormalTok{(v\_names\_states}\SpecialCharTok{ == }\StringTok{"S1"}\NormalTok{)}\SpecialCharTok{:}
\NormalTok{                                                        (n\_tunnel\_size }\SpecialCharTok{+}\DecValTok{ 1}\NormalTok{)]), }
                             \AttributeTok{S2 =}\NormalTok{ m\_M\_tunnels\_SoC[, }\StringTok{"S2"}\NormalTok{],}
                             \AttributeTok{D =}\NormalTok{ m\_M\_tunnels\_SoC[, }\StringTok{"D"}\NormalTok{])}
\end{Highlighting}
\end{Shaded}

\hypertarget{economic-outcomes}{%
\section{Economic outcomes}\label{economic-outcomes}}

In a CEA, the main outcomes are typically the total expected discounted QALYs and costs accrued by the cohort over the predefined time horizon. Below, we calculate economic outcomes from state and transition rewards. A ``state reward'' refers to a value (e.g., cost, utility) assigned to individuals for remaining in a given health state for one cycle. A ``transition reward'' refers to the increase or decrease in either costs or utilities of transitioning from one health state to another, which may be associated with a one-time cost or utility impact. The introductory tutorial describes how to incorporate state rewards in CEA in detail.\textsuperscript{\protect\hyperlink{ref-Alarid-Escudero2022b}{7}} Here, we describe and illustrate how to implement state and transition rewards together using a transition array.

\hypertarget{transition-rewards}{%
\subsection{Transition rewards}\label{transition-rewards}}

Dying (i.e., transitioning to D) incurs a one-time cost of \$2,000 that reflects the acute care that might be received immediately preceding death. We also assume a disutility and a cost increment on the transition from H to S1 (Table \ref{tab:param-table}). Incorporating transition rewards requires keeping track of the proportion of the cohort that transitions between health states in each cycle while capturing the origin and destination states for each transition. The cohort trace, \(M\), does not capture this information. However, obtaining this information is relatively straightforward in a cSTM and is described in detail by Krijkamp et al.~(2020).\textsuperscript{\protect\hyperlink{ref-Krijkamp2019}{9}} Briefly, this approach involves changing the core computation in a traditional cSTM, from \(m_t P_t\) to \(\text{diag}(m_t) P_t\). This simple change allows us to compute the proportion of the cohort that transitions between any two states in a cycle \(t\). The result is no longer a cohort trace matrix but rather a three-dimensional array that we refer to as a transition-dynamics array (\(\mathbf{A}\)) with dimensions \(n_S \times n_S \times [n_T+1]\). The \(t-\)th slice of \(\mathbf{A}\), \(A_t\), is the matrix that stores the proportion of the population that transitioned between any two states from cycles \(t-1\) to \(t\). Similarly, we define the transition rewards by the states of origin and destination.

To account for state and transition rewards, we create a \emph{matrix} of rewards \(R_t\) of dimensions \(n_S \times n_S\). The off-diagonal entries of \(R_t\) store the transition rewards, and the diagonal of \(R_t\) stores the state rewards for cycle \(t\) and assumes that rewards occur at the beginning of the cycle.\textsuperscript{\protect\hyperlink{ref-Krijkamp2019}{9}} The array of rewards, \(\mathbf{R}\), contain all \(R_t\) for all cycles. 
Finally, we multiply this matrix by \(A_t\), the \(t\)-th slice of \(A\), apply discounting, within-cycle correction, and compute the overall reward for each strategy outcome. Below, we illustrate these computations in R.

To compute \(\mathbf{A}\) for the simulation-time-dependent Sick-Sicker model under SoC, we initialize a three-dimensional array \texttt{a\_A\_SoC} of dimensions \(n_S \times n_S \times [n_T+1]\) and set the diagonal of the first slice to the initial state vector \texttt{v\_m\_init}.

\begin{Shaded}
\begin{Highlighting}[]
\CommentTok{\# Initialize transition{-}dynamics array under SoC}
\NormalTok{a\_A\_SoC }\OtherTok{\textless{}{-}} \FunctionTok{array}\NormalTok{(}\DecValTok{0}\NormalTok{,}
             \AttributeTok{dim =} \FunctionTok{c}\NormalTok{(n\_states, n\_states, (n\_cycles }\SpecialCharTok{+} \DecValTok{1}\NormalTok{)),}
             \AttributeTok{dimnames =} \FunctionTok{list}\NormalTok{(v\_names\_states, v\_names\_states, }\DecValTok{0}\SpecialCharTok{:}\NormalTok{n\_cycles))}
\CommentTok{\# Set first slice to the initial state vector in its diagonal}
\FunctionTok{diag}\NormalTok{(a\_A\_SoC[, , }\DecValTok{1}\NormalTok{]) }\OtherTok{\textless{}{-}}\NormalTok{ v\_m\_init}
\end{Highlighting}
\end{Shaded}

We then compute a matrix multiplication between a diagonal matrix of each of the \(t\)-th rows of the cohort trace matrix under SoC denoted as \texttt{diag(m\_M\_SoC{[}t,\ {]})}, by the \(t\)-th matrix of the array of the transition matrix, \texttt{a\_P\_SoC{[},\ ,\ t{]}}, over all \(n_T\) cycles.

\begin{Shaded}
\begin{Highlighting}[]
\CommentTok{\# Iterative solution to produce the transition{-}dynamics array under SoC}
\ControlFlowTok{for}\NormalTok{ (t }\ControlFlowTok{in} \DecValTok{1}\SpecialCharTok{:}\NormalTok{n\_cycles)\{}
\NormalTok{  a\_A\_SoC[, , t }\SpecialCharTok{+} \DecValTok{1}\NormalTok{] }\OtherTok{\textless{}{-}} \FunctionTok{diag}\NormalTok{(m\_M\_SoC[t, ]) }\SpecialCharTok{\%*\%}\NormalTok{ a\_P\_SoC[, , t]}
\NormalTok{\}}
\end{Highlighting}
\end{Shaded}

The arrays of rewards for costs and utilities for the simulation-time-dependent Sick-Sicker cSTM under SoC are created by filling each row across the third dimension with the vector of state rewards.

\begin{Shaded}
\begin{Highlighting}[]
\CommentTok{\# Arrays of state and transition rewards under SoC}
\CommentTok{\# Utilities}
\NormalTok{a\_R\_u\_SoC }\OtherTok{\textless{}{-}} \FunctionTok{array}\NormalTok{(}\FunctionTok{matrix}\NormalTok{(v\_u\_SoC, }\AttributeTok{nrow =}\NormalTok{ n\_states, }\AttributeTok{ncol =}\NormalTok{ n\_states, }\AttributeTok{byrow =}\NormalTok{ T), }
                  \AttributeTok{dim =} \FunctionTok{c}\NormalTok{(n\_states, n\_states, n\_cycles }\SpecialCharTok{+} \DecValTok{1}\NormalTok{),}
                  \AttributeTok{dimnames =} \FunctionTok{list}\NormalTok{(v\_names\_states, v\_names\_states, }\DecValTok{0}\SpecialCharTok{:}\NormalTok{n\_cycles))}
\CommentTok{\# Costs}
\NormalTok{a\_R\_c\_SoC }\OtherTok{\textless{}{-}} \FunctionTok{array}\NormalTok{(}\FunctionTok{matrix}\NormalTok{(v\_c\_SoC, }\AttributeTok{nrow =}\NormalTok{ n\_states, }\AttributeTok{ncol =}\NormalTok{ n\_states, }\AttributeTok{byrow =}\NormalTok{ T), }
                  \AttributeTok{dim =} \FunctionTok{c}\NormalTok{(n\_states, n\_states, n\_cycles }\SpecialCharTok{+} \DecValTok{1}\NormalTok{),}
                  \AttributeTok{dimnames =} \FunctionTok{list}\NormalTok{(v\_names\_states, v\_names\_states, }\DecValTok{0}\SpecialCharTok{:}\NormalTok{n\_cycles))}
\end{Highlighting}
\end{Shaded}

To account for the transition rewards, we either add or subtract them in the corresponding location of the reward matrix representing the transitions of interest. Thus, for example, to account for the disutility of transitioning from H to S1 under SoC, we subtract the disutility to the entry of the array of rewards corresponding to the transition from H to S1 across all cycles.

\begin{Shaded}
\begin{Highlighting}[]
\CommentTok{\# Add disutility due to transition from H to S1}
\NormalTok{a\_R\_u\_SoC[}\StringTok{"H"}\NormalTok{, }\StringTok{"S1"}\NormalTok{, ] }\OtherTok{\textless{}{-}}\NormalTok{ a\_R\_u\_SoC[}\StringTok{"H"}\NormalTok{, }\StringTok{"S1"}\NormalTok{, ] }\SpecialCharTok{{-}}\NormalTok{ du\_HS1}
\end{Highlighting}
\end{Shaded}

In a similar approach, we add the costs of transitioning from H to S1 and the cost of dying.

\begin{Shaded}
\begin{Highlighting}[]
\CommentTok{\# Add transition cost due to transition from H to S1}
\NormalTok{a\_R\_c\_SoC[}\StringTok{"H"}\NormalTok{, }\StringTok{"S1"}\NormalTok{, ] }\OtherTok{\textless{}{-}}\NormalTok{ a\_R\_c\_SoC[}\StringTok{"H"}\NormalTok{, }\StringTok{"S1"}\NormalTok{, ] }\SpecialCharTok{+}\NormalTok{ ic\_HS1}
\CommentTok{\# Add transition cost of dying from all non{-}dead states}
\NormalTok{a\_R\_c\_SoC[}\SpecialCharTok{{-}}\NormalTok{n\_states, }\StringTok{"D"}\NormalTok{, ] }\OtherTok{\textless{}{-}}\NormalTok{ a\_R\_c\_SoC[}\SpecialCharTok{{-}}\NormalTok{n\_states, }\StringTok{"D"}\NormalTok{, ] }\SpecialCharTok{+}\NormalTok{ ic\_D}
\NormalTok{a\_R\_c\_SoC[, , }\DecValTok{1}\NormalTok{]}
\end{Highlighting}
\end{Shaded}

\begin{verbatim}
##       H    S1    S2    D
## H  2000 17000 27000 2000
## S1 2000 16000 27000 2000
## S2 2000 16000 27000 2000
## D  2000 16000 27000    0
\end{verbatim}

The state and transition rewards are applied to the model dynamics by element-wise multiplication between \(\mathbf{A}\) and \(\mathbf{R}\), indicated by the \(\odot\) sign, which produces the array of economic outcomes for all \(n_T\) cycles, \(\mathbf{Y}\). Formally,
\begin{equation}
  \mathbf{Y} = \mathbf{A} \odot \mathbf{R}
  \label{eq:array-outputs}
\end{equation}

To obtain \(\mathbf{Y}\) for QALYs and costs for all four strategies, we apply Equation \eqref{eq:array-outputs} by the element-wise multiplication of the transition array \texttt{a\_A\_SoC} by the corresponding array of rewards.

\begin{Shaded}
\begin{Highlighting}[]
\CommentTok{\# For SoC}
\NormalTok{a\_Y\_c\_SoC }\OtherTok{\textless{}{-}}\NormalTok{ a\_A\_SoC }\SpecialCharTok{*}\NormalTok{ a\_R\_c\_SoC}
\NormalTok{a\_Y\_u\_SoC }\OtherTok{\textless{}{-}}\NormalTok{ a\_A\_SoC }\SpecialCharTok{*}\NormalTok{ a\_R\_u\_SoC}
\end{Highlighting}
\end{Shaded}

The total rewards for each health state at cycle \(t\), \(\mathbf{y}_t\), is obtained by summing the rewards across all \(j = 1,\ldots, n_S\) health states for all \(n_T\) cycles.
\begin{equation}
  \mathbf{y}_t = \mathbf{1}^T Y_t = \left[\sum_{i=1}^{n_S}{Y_{[i,1,t]}}, \sum_{i=1}^{n_S}{Y_{[i,2,t]}}, \dots , \sum_{i=1}^{n_S}{Y_{[i,n_S,t]}}\right].
  \label{eq:exp-rewd-trans}
\end{equation}

To obtain the expected costs and QALYs per cycle under SoC, \(\mathbf{y}\), we apply Equation \eqref{eq:exp-rewd-trans} again across all the matrices of the third dimension of \(\mathbf{Y}\) for all the outcomes. The complete code for all the strategies is presented in the Supplementary Material.

\begin{Shaded}
\begin{Highlighting}[]
\CommentTok{\# Vectors of rewards under SoC}
\CommentTok{\# QALYs}
\NormalTok{v\_qaly\_SoC }\OtherTok{\textless{}{-}} \FunctionTok{rowSums}\NormalTok{(}\FunctionTok{t}\NormalTok{(}\FunctionTok{colSums}\NormalTok{(a\_Y\_u\_SoC)))}
\CommentTok{\# Costs}
\NormalTok{v\_cost\_SoC }\OtherTok{\textless{}{-}} \FunctionTok{rowSums}\NormalTok{(}\FunctionTok{t}\NormalTok{(}\FunctionTok{colSums}\NormalTok{(a\_Y\_c\_SoC)))}
\end{Highlighting}
\end{Shaded}

\hypertarget{within-cycle-correction-and-discounting-future-rewards}{%
\subsubsection{Within-cycle correction and discounting future rewards}\label{within-cycle-correction-and-discounting-future-rewards}}
We use Simpson's 1/3rd rule for within-cycle correction (WCC),\textsuperscript{\protect\hyperlink{ref-Elbasha2016a}{18}} and use exponential discounting for costs and QALYs. Please see the Supplementary Material and introductory cSTM tutorial for a detailed description of implementing discounting and WCC in R.\textsuperscript{\protect\hyperlink{ref-Alarid-Escudero2022b}{7}}

\hypertarget{incremental-cost-effectiveness-ratios-icers}{%
\section{Incremental cost-effectiveness ratios (ICERs)}\label{incremental-cost-effectiveness-ratios-icers}}

We followed the coding approach described in the introductory cSTM tutorial to conduct the cost-effectiveness analysis.\textsuperscript{\protect\hyperlink{ref-Alarid-Escudero2022b}{7}} We use the R package \texttt{dampack} (\url{https://cran.r-project.org/web/packages/dampack/})\textsuperscript{\protect\hyperlink{ref-Alarid-Escudero2021}{19}} to calculate the incremental costs and effectiveness and the incremental cost-effectiveness ratio (ICER) between non-dominated strategies. \texttt{dampack} organizes and formats the results as a data frame, \texttt{df\_cea}, that can be printed as a formatted table.

\begin{figure}[H]

{\centering \includegraphics{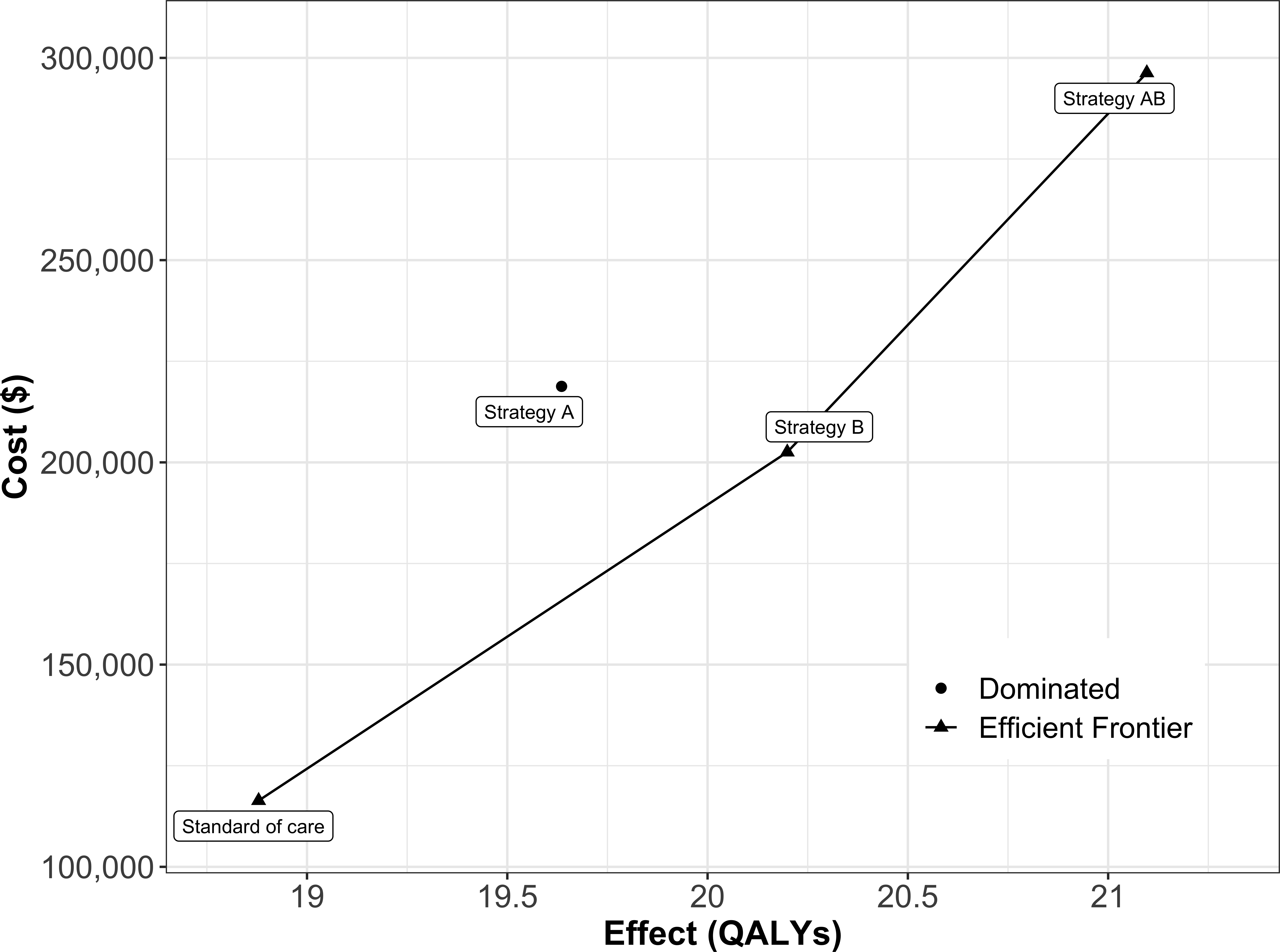} 

}

\caption{Cost-effectiveness efficient frontier of all four strategies for the simulation-time-dependent Sick-Sicker model.}\label{fig:Sick-Sicker-CEA-AgeDep}
\end{figure}

The SoC Strategy is the least costly and effective strategy, followed by Strategy B, producing an expected benefit of 1.32 QALYs per individual for an additional expected cost of \$86,162 with an ICER of \$65,288/QALY followed by Strategy AB with an ICER \$104,461/QALY. Strategy A is a dominated strategy. The results of the CEA of the simulation-time-dependent Sick-Sicker model are presented in Table \ref{tab:table-cea}. The non-dominated strategies, SoC, B, and AB, form the cost-effectiveness efficient frontier of the CEA based on the simulation-time-dependent Sick-Sicker model (Figure \ref{fig:Sick-Sicker-CEA-AgeDep}).

\begin{table}[!h]

\caption{\label{tab:table-cea}Cost-effectiveness analysis results for the simulation-time-dependent Sick-Sicker model. ND: Non-dominated strategy; D: Dominated strategy.}
\centering
\begin{tabular}[t]{rcccccc}
\toprule{}
Strategy & Costs (\$) & QALYs & Incremental Costs (\$) & Incremental QALYs & ICER (\$/QALY) & Status\\
\midrule{}
Standard of care & 116,374 & 18.879 & NA & NA & NA & ND\\
Strategy B & 202,536 & 20.199 & 86,162 & 1.320 & 65,288 & ND\\
Strategy AB & 296,300 & 21.097 & 93,764 & 0.898 & 104,461 & ND\\
\midrule
Strategy A & 218,789 & 19.636 & NA & NA & NA & D\\
\bottomrule{}
\end{tabular}
\end{table}

\hypertarget{probabilistic-sensitivity-analysis}{%
\section{Probabilistic sensitivity analysis}\label{probabilistic-sensitivity-analysis}}

We conducted a probabilistic sensitivity analysis (PSA) to quantify the effect of model parameter uncertainty on cost-effectiveness outcomes.\textsuperscript{\protect\hyperlink{ref-Briggs2012}{20}} In a PSA, we randomly draw parameter sets from distributions that reflect the current uncertainty in model parameter estimates. The parameters' distributions and their values are described in Table \ref{tab:param-table} and in more detail in the Supplementary Material. We compute model outcomes for each sampled set of parameter values (e.g., total discounted cost and QALYs) for each strategy. We follow the steps to conduct a PSA from a previously published article describing the Decision Analysis in R for technologies in Health (DARTH) coding framework.\textsuperscript{\protect\hyperlink{ref-Alarid-Escudero2019e}{14}} The R code to generate the PSA is included in the Supplementary Material. The description of the functions and steps and a brief discussion on choosing distributions are described in more detail in the introductory tutorial.\textsuperscript{\protect\hyperlink{ref-Alarid-Escudero2022b}{7}}

\begin{figure}[H]

{\centering \includegraphics{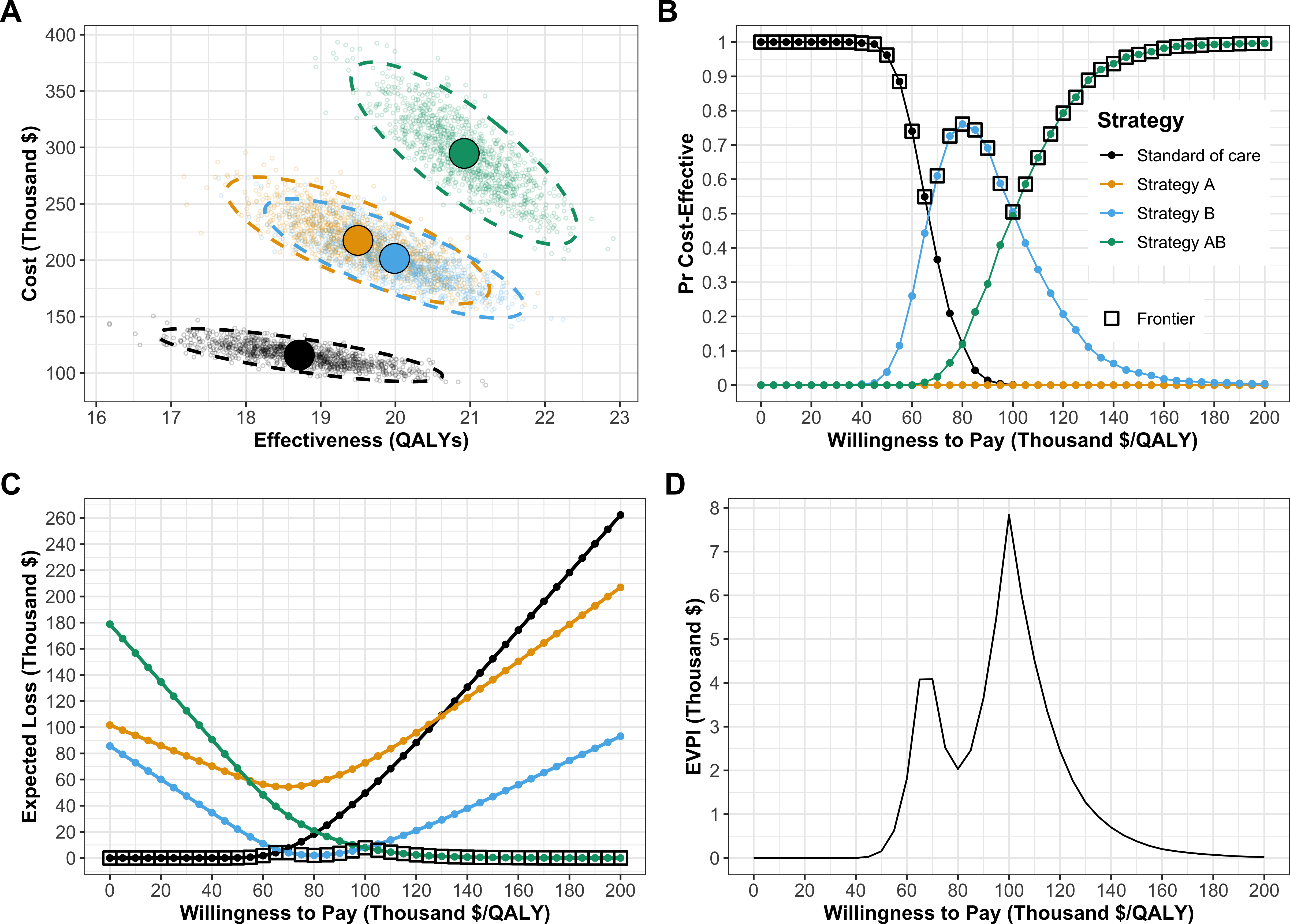} 

}

\caption{Figures generated from the probabilistic sensitivity analysis (PSA) output. A) Cost-effectiveness scatter plot. B) Cost-effectiveness acceptability curves (CEACs) and frontier (CEAF). C) Expected loss curves (ELCs). D) Expected value of perfect information (EVPI).}\label{fig:PSA-figures}
\end{figure}

To conduct the PSA of the CEA using the simulation time-dependent Sick-Sicker cSTM, we sampled 1,000 parameter sets. We computed the total discounted costs and QALYs of each simulated strategy for each parameter set. We generated the cost-effectiveness scatter plot,\textsuperscript{\protect\hyperlink{ref-Briggs2002}{21}} where each of the 4,000 simulations (i.e., 1,000 combinations of total discounted expected costs and QALYs for each of the four strategies) are plotted as a point in the graph (Figure \ref{fig:PSA-figures}A)). We also added the 95\% confidence ellipse, and the expected values of the total discounted costs and QALYs for each strategy. The CE scatter plot shows that Strategy AB has the highest expected costs and QALYs. Standard of care has the lowest expected cost and QALYs. Strategy B is more effective and least costly than Strategy A. Therefore, Strategy A is strongly dominated by Strategy B.

In Figure \ref{fig:PSA-figures}B, we present the cost-effectiveness acceptability curves (CEACs), showing the probability that each strategy is cost-effective, and the cost-effectiveness frontier (CEAF), displaying the probability of the optimal strategy being cost-effective over a range of willingness-to-pay (WTP) thresholds. At WTP thresholds less than \$70,000 per QALY, SoC is the strategy with the highest probability of being cost-effective and the highest expected NMB. Strategy B has the highest probability of being cost-effective and the highest expected NMB for WTP thresholds greater than \$70,000 and lower than \$105,000 per QALY. Strategy AB has the highest expected NMB for WTP thresholds greater than or equal to \$105,000 and is the strategy with the highest probability of being cost-effective.

We quantify the expected loss from each strategy over a range of WTP thresholds with the expected loss curves (ELCs) to complement the PSA results (Figure \ref{fig:PSA-figures}C). The expected loss considers both the probability of making the wrong decision and the magnitude of the loss due to this decision, representing the foregone benefits of choosing a suboptimal strategy. The SoC Strategy has the lowest expected loss for WTP thresholds less than \$65,000 per QALY, Strategy B has the lowest expected loss for WTP thresholds greater than or equal to \$70,000 and less than \$105,000. Strategy AB has the lowest expected loss for WTP thresholds greater than or equal \$105,000 per QALY. At a WTP threshold of \$100,000 per QALY, the EVPI is highest at \$7,836 (Figures \ref{fig:PSA-figures}C-D). For a more detailed description of these outputs and the R code to generate them, we refer the reader to a previous publication by our group.\textsuperscript{\protect\hyperlink{ref-Alarid-Escudero2019}{22}}

\hypertarget{epidemiological-outcomes}{%
\section{Epidemiological outcomes}\label{epidemiological-outcomes}}

cSTMs can be used to generate epidemiological outcomes that can be of direct interest to the decision maker or be helpful for model calibration and validation. Some common epidemiological outcomes include survival, prevalence, incidence, the average number of events, and lifetime risk of events.\textsuperscript{\protect\hyperlink{ref-Siebert2012c}{23}} We provide the epidemiological definitions of some of these outcomes and how they can be generated from the trace and transition probability objects of a cSTM using the simulation time-dependent Sick-Sicker cSTM under SoC. Similar to the PSA of the economic outcomes, we generated the epidemiological outcomes for each of the parameter sets drawn from their distributions and applied the equations described below. We summarized the outcomes with the 95\% posterior predicted interval (PI), defined as the estimated range between the 2.5th and 97.5th percentiles of the model-predicted posterior outputs. The Supplemental Material provides the code to generate these outcomes from the state-residence dependent cSTM.

\hypertarget{survival-probability}{%
\subsection{Survival probability}\label{survival-probability}}

\begin{figure}[H]

{\centering \includegraphics{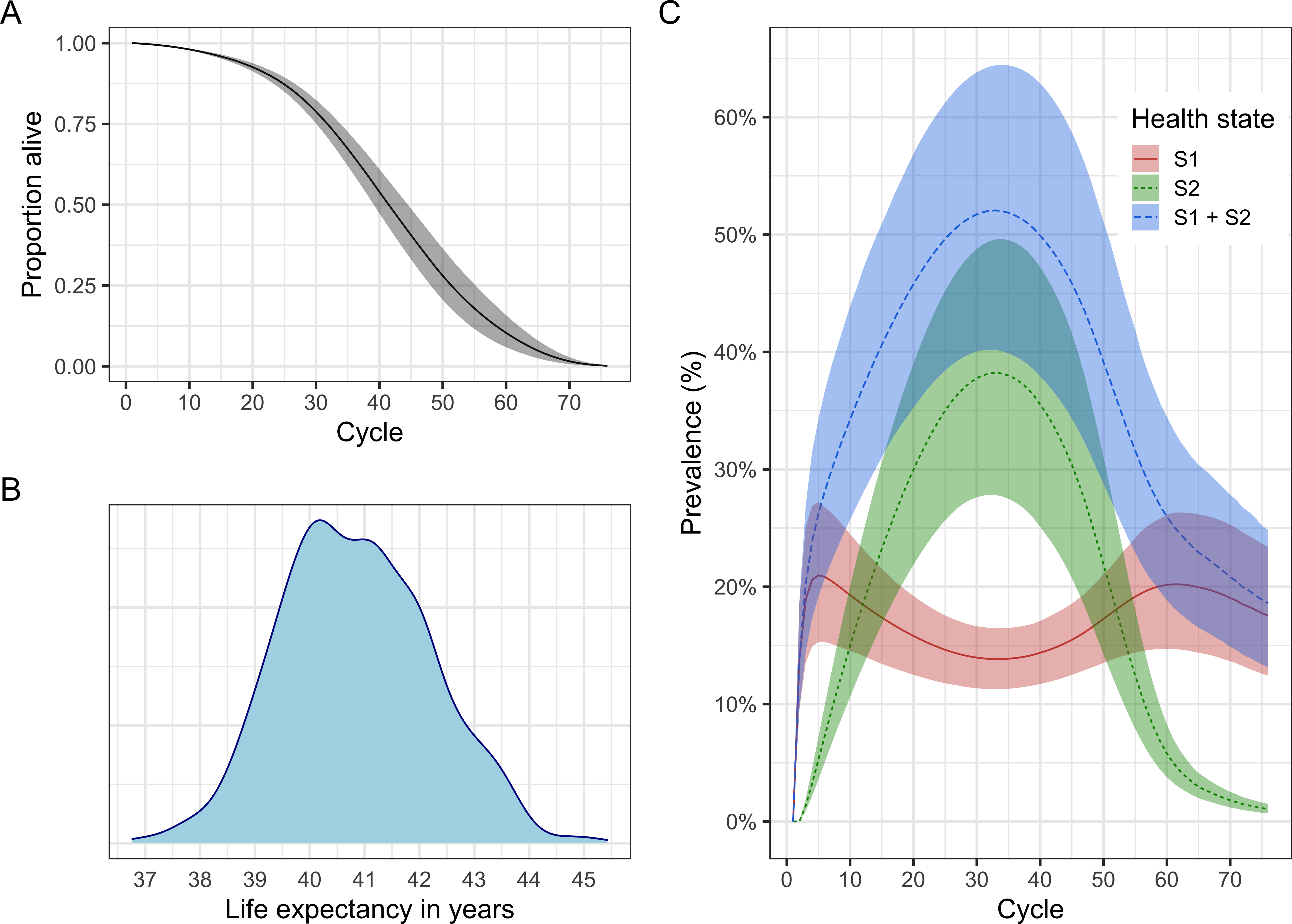} 

}

\caption{Epidemiological outcomes generated from the probabilistic sensitivity analysis (PSA) output for the simulation-time dependent cSTM. A) Survival curve. B) Posterior density of life expectancy. C) Prevalence of sick states over time. The shaded area in A and C shows the 95\% posterior model-predictive interval of the outcomes and colored lines shows the posterior model-predicted mean based on 1,000 simulations.}\label{fig:PSA-EPI-figures}
\end{figure}

The survival probability, \(S(t)\), captures the proportion of the cohort remaining alive by cycle \(t\). To estimate \(S(t)\) from the simulated cohort of the simulation-time-dependent Sick-Sicker model, shown in Figure \ref{fig:PSA-EPI-figures}A, we sum the proportions of the non-death states for all \(n_T\) cycles in \texttt{m\_M\_SoC}.

\begin{Shaded}
\begin{Highlighting}[]
\NormalTok{v\_S\_SoC }\OtherTok{\textless{}{-}} \FunctionTok{rowSums}\NormalTok{(m\_M\_SoC[, }\SpecialCharTok{{-}}\FunctionTok{which}\NormalTok{(v\_names\_states }\SpecialCharTok{==} \StringTok{"D"}\NormalTok{)]) }\CommentTok{\# vector with survival curve}
\end{Highlighting}
\end{Shaded}

\hypertarget{life-expectancy}{%
\subsection{Life expectancy}\label{life-expectancy}}

Life expectancy (LE) refers to the expected number of time units remaining to be alive.\textsuperscript{\protect\hyperlink{ref-Lee2003a}{24}} In continuous-time, LE is the area under the entire survival curve.\textsuperscript{\protect\hyperlink{ref-Klein2003}{25}}

\[
LE = \int_{t=0}^{\infty}{S(t) dt}.
\]

In discrete-time using cSTMs, we often calculate restricted LE over a fixed time horizon (e.g., \(n_T\)) at which most of the cohort has transitioned to the Dead state and is defined as

\[
  LE = \sum_{t=0}^{n_T}{S(t)}.
\]

In the simulation-time dependent Sick-Sicker model, where we simulated a cohort over \(n_T\)= 75 cycles, restricted life expectancy \texttt{le\_SoC} is 40.8 (95\% credible interval:{[}38.4, 43.6{]}) cycles (Figure \ref{fig:PSA-EPI-figures}B), which is calculated as

\begin{Shaded}
\begin{Highlighting}[]
\NormalTok{le\_SoC }\OtherTok{\textless{}{-}} \FunctionTok{sum}\NormalTok{(v\_S\_SoC) }\CommentTok{\# life expectancy}
\end{Highlighting}
\end{Shaded}

Note that this equation expresses LE in the units of \(t\). We use an annual cycle length; thus, the resulting LE will be in years. Analysts can also use other cycle lengths (e.g., monthly or daily), but the LE must be correctly converted to the desired unit if different than the cycle length units.

\hypertarget{prevalence}{%
\subsection{Prevalence}\label{prevalence}}

Prevalence is defined as the proportion of the population or cohort with a specific condition (or being in a particular health state) among those alive.\textsuperscript{\protect\hyperlink{ref-Rothman2008h}{26}} To calculate the prevalence of S1 at cycle \(t\), \(\text{prev}(t)_i\), we compute the ratio between the proportion of the cohort in S1 and the proportion alive at that cycle.\textsuperscript{\protect\hyperlink{ref-Keiding1991}{27}} The proportion of the cohort alive is given by the survival probability \(S(t)\) defined above. The individual prevalence of the S1 and S2 health states and the overall prevalence of sick individuals (i.e., S1 + S2) of the age-dependent Sick-Sicker cSTM at each cycle \(t\) is computed as follows and are shown in Figure \ref{fig:PSA-EPI-figures}C.

\begin{Shaded}
\begin{Highlighting}[]
\NormalTok{v\_prev\_S1\_SoC   }\OtherTok{\textless{}{-}}\NormalTok{ m\_M\_SoC[, }\StringTok{"S1"}\NormalTok{] }\SpecialCharTok{/}\NormalTok{ v\_S\_SoC          }\CommentTok{\# vector with prevalence of Sick}
\NormalTok{v\_prev\_S2\_SoC   }\OtherTok{\textless{}{-}}\NormalTok{ m\_M\_SoC[, }\StringTok{"S2"}\NormalTok{] }\SpecialCharTok{/}\NormalTok{ v\_S\_SoC          }\CommentTok{\# vector with prevalence of Sicker}
\NormalTok{v\_prev\_S1S2\_SoC }\OtherTok{\textless{}{-}} \FunctionTok{rowSums}\NormalTok{(m\_M\_SoC[, }\FunctionTok{c}\NormalTok{(}\StringTok{"S1"}\NormalTok{, }\StringTok{"S2"}\NormalTok{)])}\SpecialCharTok{/}\NormalTok{v\_S\_SoC }\CommentTok{\# prevalence of Sick and Sicker}
\end{Highlighting}
\end{Shaded}

\hypertarget{discussion}{%
\section{Discussion}\label{discussion}}

In this tutorial, we conceptualize time-dependent cSTMs with their mathematical description and a walk-through of their implementation for a CEA in R using the Sick-Sicker example. We described two types of time dependence: on the time since the start of the simulation (simulation-time dependence) or on the time spent in a health state (state-residence time dependence). We illustrated how to generate various epidemiological outcomes from the model, incorporate transition rewards in CEAs, and conduct a PSA.

We implemented simulation-time dependence by expanding the two-dimensional transition probability matrix into a three-dimensional transition probability array, where the third dimension captures time since the start of the simulation. However, there are alternative implementations of simulation-time dependence in cSTMs. For example, the model could be coded such that the time-varying elements of the transition probability matrix \(P_t\) are updated at each time point \(t\) as the simulation is run. This would eliminate the need for the transition probability array \texttt{a\_P}, reducing computer memory requirements. But this comes at the expense of increasing the number of operations at every cycle, potentially slowing down the simulation.

We incorporated state-residence time dependence using tunnel states by expanding the corresponding health states on the first and second dimensions of the three-dimensional array to account for time spent in the current state in addition to simulation-time dependence. Another approach to account for state-residence time dependence is to use a 3-dimensional transition probability array with dimensions for the current state, future state, and time in the current state.\textsuperscript{\protect\hyperlink{ref-Hawkins2005}{28}} However, combining simulation-time and state-residence time dependencies could necessitate a four-dimensional array, which may prove challenging.

Any time-varying feature in a discrete-time model can most generally be implemented as tunnel states with, at the extreme, every state having a different tunnel state for each time step. The cohort would then progressively move through these tunnel states to capture their progression through time and the model features (e.g., transition probabilities, costs, or utilities) that change over time. Using time-varying transition probabilities is a shortcut when the cohort experiences these time-varying processes simultaneously as a function of the time from the simulation start. Even if the time-varying process has a different periodicity than the cycle length, either tunnel states or time-varying transition probabilities can be used to capture these effects. However, this time-varying process must be represented or approximated over an integer number of cycle lengths.

As described in the introductory tutorial, cSTMs are recommended when the number of states is considered ``not too large.''\textsuperscript{\protect\hyperlink{ref-Siebert2012c}{23}} Incorporating time dependence in cSTMs using the provided approaches requires expanding the number of states by creating a multidimensional array for simulation-time dependence and/or creating tunnel states for state-residence time dependence, increasing the amount of computer memory (RAM) required. For example, as state-residence time dependence extends in more health states, the total number of health states will increase, causing a ``state explosion'`. It is possible to build reasonably complex time-dependent cSTMs in R as long as there is sufficient RAM to store the transition probability array and outputs of interest. For example, a typical PC with 8GB of RAM can handle a transition probability array of about 1,000 states and 600 time-cycle slices. However, if a high degree of granularity is desired, the dimensions of these data objects can grow quickly; if the required number of states gets too large and difficult to code, it may be preferable to use a stochastic (Monte Carlo) version of the state-transition model -- often called individual-based state-transition models (iSTM) or microsimulation models -- rather than a cohort simulation model.\textsuperscript{\protect\hyperlink{ref-Siebert2012c}{23}} In an iSTM, the risks and rewards of simulated individuals need not depend only on a current health state; they may also depend on an individual's characteristics and attributes. In addition, modelers can store health state history and other events over time for each individual to determine the risk of new events and corresponding costs and effects. Thus, we recommend carefully considering the required model structure before implementing it. An iSTM will require additional functions to describe the dependence of transition probabilities and rewards on individuals' histories. A previous tutorial showed how to construct these functions for an iSTM using the Sick-Sicker example.\textsuperscript{\protect\hyperlink{ref-Krijkamp2018}{29}}

We described the concept and implementation of transition rewards. While these event-driven impacts could instead be captured by expanding the model state-space to include an initial disease state or a pre-death, end-of-life state, we can avoid state-space expansion by calculating incurred rewards based on the proportion of the cohort making the relevant transition. For implementation, this requires storing not just the cohort trace, which reflects how many individuals are in each state at a given cycle, but also a cohort state-transition array, which records how many individuals are making each possible transition at a given cycle.\textsuperscript{\protect\hyperlink{ref-Krijkamp2019}{9}} We model the effectiveness of Treatment B as a hazard ratio. However, other effect measures can be used, such as odds ratio, risk ratio, or risk differences depending on the methods used to estimate them.\textsuperscript{\protect\hyperlink{ref-Kuntz2001}{30}--\protect\hyperlink{ref-Gidwani2020}{32}} We assume that all parameters are known for the case study. However, in a real-world CEA, parameters are estimated from electronic health records,\textsuperscript{\protect\hyperlink{ref-Rodriguez2021}{33}} clinical trials, and observational studies, and when there is no individual-level data to inform them, through model calibration.\textsuperscript{\protect\hyperlink{ref-Welton2005}{34}}

We use \texttt{base} R to implement time-dependent cSTMs as the most transparent way to illustrate how to translate the mathematical specification of cSTMs into code. Using a \texttt{base} R implementation maximizes flexibility and customizability to model any disease that can be specified as a cSTM. However, there are R packages for building and running cSTMs and conducting CEAs, such as \texttt{heemod}\textsuperscript{\protect\hyperlink{ref-Filipovic-Pierucci2017}{35}} and \texttt{hesim},\textsuperscript{\protect\hyperlink{ref-Incerti2021}{36}} that are made such that they can support modeling arbitrary patient populations or many treatment strategies. Since much of the programming burden has already been done, these packages can reduce programming time. Users don't need to write functions themselves but instead need to call a series of prewritten functions. And in the case of \texttt{hesim}, all the memory pre-allocation and loops are made at the \texttt{C++} level, making it more efficient than looping over \texttt{base} R but might require more RAM. Ultimately, it is up to the analyst to decide the most appropriate approach to implementing their cSTM-based CEAs.

This tutorial extends our conceptualization of time-independent cSTMs to allow for time dependence. It provides a step-by-step guide to implementing time-dependent cSTMs in R to generate epidemiological and economic outcomes, accounts for transition rewards, and conducts a CEA and a corresponding PSA. We hope that health decision scientists and health economists find this tutorial helpful in developing their cSTMs in a more flexible, efficient, and open-source manner. Ultimately, our goal is to facilitate R in health economic evaluations with the overall aim to increase model transparency and reproducibility.

\section*{Acknowledgements}\label{acknowledgements}
\addcontentsline{toc}{section}{Acknowledgements}

Dr. Alarid-Escudero was supported by grants U01-CA199335, U01-CA253913, U01-CA265750, and U01-CA265729 from the National Cancer Institute (NCI) as part of the Cancer Intervention and Surveillance Modeling Network (CISNET), and a grant by the Gordon and Betty Moore Foundation. Miss Krijkamp was supported by the Society for Medical Decision Making (SMDM) fellowship through a grant by the Gordon and Betty Moore Foundation (GBMF7853). Dr. Enns was supported by a grant from the National Institute of Allergy and Infectious Diseases of the National Institutes of Health under award no. K25AI118476. Dr. Hunink received research funding from the American Diabetes Association, the Netherlands Organization for Health Research and Development, the German Innovation Fund, Netherlands Educational Grant (``Studie Voorschot Middelen''), and the Gordon and Betty Moore Foundation. Dr. Jalal was supported by a grant from the National Institute on Drug Abuse of the National Institute of Health under award no. K01DA048985, and U01-CA265750 from NCI as part of CISNET. The content is solely the responsibility of the authors and does not necessarily represent the official views of the National Institutes of Health. The funding agencies had no role in the design of the study, interpretation of results, or writing of the manuscript. The funding agreement ensured the authors' independence in designing the study, interpreting the data, writing, and publishing the report. We also want to thank the anonymous reviewers of \emph{Medical Decision Making} for their valuable suggestions and the students who took our classes where we refined these materials.

\hypertarget{references}{%
\section*{References}\label{references}}
\addcontentsline{toc}{section}{References}

\hypertarget{refs}{}
\begin{CSLReferences}{0}{0}
\leavevmode\hypertarget{ref-Suijkerbuijk2018}{}%
\CSLLeftMargin{1. }
\CSLRightInline{Suijkerbuijk AWM, Van Hoek AJ, Koopsen J, De Man RA, Mangen MJJ, De Melker HE, et al. {Cost-effectiveness of screening for chronic hepatitis B and C among migrant populations in a low endemic country}. PLoS ONE. 2018;13(11):1--6. }

\leavevmode\hypertarget{ref-Sathianathen2018a}{}%
\CSLLeftMargin{2. }
\CSLRightInline{Sathianathen NJ, Konety BR, Alarid-Escudero F, Lawrentschuk N, Bolton DM, Kuntz KM. {Cost-effectiveness Analysis of Active Surveillance Strategies for Men with Low-risk Prostate Cancer}. European Urology {[}Internet{]}. 2019;75(6):910--7. Available from: \url{https://linkinghub.elsevier.com/retrieve/pii/S0302283818308534}}

\leavevmode\hypertarget{ref-Lu2018b}{}%
\CSLLeftMargin{3. }
\CSLRightInline{Lu S, Yu Y, Fu S, Ren H. {Cost-effectiveness of ALK testing and first-line crizotinib therapy for non-small-cell lung cancer in China}. PLoS ONE. 2018;13(10):1--2. }

\leavevmode\hypertarget{ref-Djatche2018}{}%
\CSLLeftMargin{4. }
\CSLRightInline{Djatche LM, Varga S, Lieberthal RD. {Cost-Effectiveness of Aspirin Adherence for Secondary Prevention of Cardiovascular Events}. PharmacoEconomics - Open {[}Internet{]}. 2018;2(4):371--80. Available from: \url{https://doi.org/10.1007/s41669-018-0075-2}}

\leavevmode\hypertarget{ref-Pershing2014}{}%
\CSLLeftMargin{5. }
\CSLRightInline{Pershing S, Enns EA, Matesic B, Owens DK, Goldhaber-Fiebert JD. {Cost-Effectiveness of Treatment of Diabetic Macular Edema}. Annals of Internal Medicine. 2014;160(1):18--29. }

\leavevmode\hypertarget{ref-Smith-Spangler2010}{}%
\CSLLeftMargin{6. }
\CSLRightInline{Smith-Spangler CM, Juusola JL, Enns EA, Owens DK, Garber AM. {Population Strategies to Decrease Sodium Intake and the Burden of Cardiovascular Disease: A Cost-Effectiveness Analysis}. Annals of Internal Medicine {[}Internet{]}. 2010;152(8):481--7. Available from: \url{http://annals.org/article.aspx?articleid=745729}}

\leavevmode\hypertarget{ref-Alarid-Escudero2022b}{}%
\CSLLeftMargin{7. }
\CSLRightInline{Alarid-Escudero F, Krijkamp EM, Enns EA, Yang A, Hunink MGM, Pechlivanoglou P, et al. {An Introductory Tutorial on Cohort State-Transition Models in R Using a Cost-Effectiveness Analysis Example}. Medical Decision Making. 2022;In Press. }

\leavevmode\hypertarget{ref-Snowsill2019}{}%
\CSLLeftMargin{8. }
\CSLRightInline{Snowsill T. {A New Method for Model-Based Health Economic Evaluation Utilizing and Extending Moment-Generating Functions}. Medical Decision Making {[}Internet{]}. 2019;39(5):523--39. Available from: \url{http://journals.sagepub.com/doi/10.1177/0272989X19860119}}

\leavevmode\hypertarget{ref-Krijkamp2019}{}%
\CSLLeftMargin{9. }
\CSLRightInline{Krijkamp EM, Alarid-Escudero F, Enns E, Pechlivanoglou P, Hunink MM, Jalal H. {A Multidimensional Array Representation of State-Transition Model Dynamics}. Medical Decision Making. 2019;In Press. }

\leavevmode\hypertarget{ref-Jalal2017b}{}%
\CSLLeftMargin{10. }
\CSLRightInline{Jalal H, Pechlivanoglou P, Krijkamp E, Alarid-Escudero F, Enns EA, Hunink MGM. {An Overview of R in Health Decision Sciences}. Medical Decision Making {[}Internet{]}. 2017;37(7):735--46. Available from: \url{http://journals.sagepub.com/doi/10.1177/0272989X16686559}}

\leavevmode\hypertarget{ref-Hunink2014}{}%
\CSLLeftMargin{11. }
\CSLRightInline{Hunink MGGM, Weinstein MC, Wittenberg E, Drummond MF, Pliskin JS, Wong JB, et al. {Decision Making in Health and Medicine} {[}Internet{]}. 2nd ed. Cambridge: Cambridge University Press; 2014. Available from: \url{http://ebooks.cambridge.org/ref/id/CBO9781139506779}}

\leavevmode\hypertarget{ref-Enns2015e}{}%
\CSLLeftMargin{12. }
\CSLRightInline{Enns EA, Cipriano LE, Simons CT, Kong CY. {Identifying Best-Fitting Inputs in Health-Economic Model Calibration: A Pareto Frontier Approach}. Medical Decision Making {[}Internet{]}. 2015;35(2):170--82. Available from: \url{http://www.ncbi.nlm.nih.gov/pubmed/24799456}}

\leavevmode\hypertarget{ref-OMahony2015}{}%
\CSLLeftMargin{13. }
\CSLRightInline{O'Mahony JF, Newall AT, Rosmalen J van. {Dealing with Time in Health Economic Evaluation: Methodological Issues and Recommendations for Practice.} PharmacoEconomics {[}Internet{]}. 2015;33(12):1265--8. Available from: \url{http://www.ncbi.nlm.nih.gov/pubmed/26105525}}

\leavevmode\hypertarget{ref-Alarid-Escudero2019e}{}%
\CSLLeftMargin{14. }
\CSLRightInline{Alarid-Escudero F, Krijkamp E, Pechlivanoglou P, Jalal H, Kao S-YZ, Yang A, et al. {A Need for Change! A Coding Framework for Improving Transparency in Decision Modeling}. PharmacoEconomics {[}Internet{]}. 2019;37(11):1329--39. Available from: \url{https://doi.org/10.1007/s40273-019-00837-x}}

\leavevmode\hypertarget{ref-Arias2017}{}%
\CSLLeftMargin{15. }
\CSLRightInline{Arias E, Heron M, Xu J. {United States Life Tables, 2014}. National Vital Statistics Reports {[}Internet{]}. 2017;66(4):63. Available from: \url{https://www.cdc.gov/nchs/data/nvsr/nvsr66/nvsr66\%7B/_\%7D04.pdf}}

\leavevmode\hypertarget{ref-Diaby2014}{}%
\CSLLeftMargin{16. }
\CSLRightInline{Diaby V, Adunlin G, Montero AJ. {Survival modeling for the estimation of transition probabilities in model-based economic evaluations in the absence of individual patient data: A tutorial}. PharmacoEconomics. 2014 Feb;32(2):101--8. }

\leavevmode\hypertarget{ref-Elbasha2016}{}%
\CSLLeftMargin{17. }
\CSLRightInline{Elbasha EH, Chhatwal J. {Theoretical foundations and practical applications of within-cycle correction methods}. Medical Decision Making. 2016;36(1):115--31. }

\leavevmode\hypertarget{ref-Elbasha2016a}{}%
\CSLLeftMargin{18. }
\CSLRightInline{Elbasha EH, Chhatwal J. {Myths and misconceptions of within-cycle correction: a guide for modelers and decision makers}. PharmacoEconomics. 2016;34(1):13--22. }

\leavevmode\hypertarget{ref-Alarid-Escudero2021}{}%
\CSLLeftMargin{19. }
\CSLRightInline{Alarid-Escudero F, Knowlton G, Easterly CA, Enns EA. Decision analytic modeling package (dampack) {[}Internet{]}. 2021. Available from: \url{https://cran.r-project.org/web/packages/dampack/\%20https://github.com/DARTH-git/dampack}}

\leavevmode\hypertarget{ref-Briggs2012}{}%
\CSLLeftMargin{20. }
\CSLRightInline{Briggs AH, Weinstein MC, Fenwick EAL, Karnon J, Sculpher MJ, Paltiel AD. {Model Parameter Estimation and Uncertainty Analysis: A Report of the ISPOR-SMDM Modeling Good Research Practices Task Force Working Group-6.} Medical Decision Making. 2012 Sep;32(5):722--32. }

\leavevmode\hypertarget{ref-Briggs2002}{}%
\CSLLeftMargin{21. }
\CSLRightInline{Briggs AH, Goeree R, Blackhouse G, O'Brien BJ. {Probabilistic Analysis of Cost-Effectiveness Models: Choosing between Treatment Strategies for Gastroesophageal Reflux Disease}. Medical Decision Making {[}Internet{]}. 2002 Jul;22(4):290--308. Available from: \url{http://mdm.sagepub.com/cgi/doi/10.1177/027298902400448867}}

\leavevmode\hypertarget{ref-Alarid-Escudero2019}{}%
\CSLLeftMargin{22. }
\CSLRightInline{Alarid-Escudero F, Enns EA, Kuntz KM, Michaud TL, Jalal H. {"Time Traveling Is Just Too Dangerous" But Some Methods Are Worth Revisiting: The Advantages of Expected Loss Curves Over Cost-Effectiveness Acceptability Curves and Frontier}. Value in Health. 2019;22(5):611--8. }

\leavevmode\hypertarget{ref-Siebert2012c}{}%
\CSLLeftMargin{23. }
\CSLRightInline{Siebert U, Alagoz O, Bayoumi AM, Jahn B, Owens DK, Cohen DJ, et al. {State-Transition Modeling: A Report of the ISPOR-SMDM Modeling Good Research Practices Task Force-3}. Medical Decision Making {[}Internet{]}. 2012;32(5):690--700. Available from: \url{http://mdm.sagepub.com/cgi/doi/10.1177/0272989X12455463}}

\leavevmode\hypertarget{ref-Lee2003a}{}%
\CSLLeftMargin{24. }
\CSLRightInline{Lee ET, Wang JW. {Statistical methods for Survival Data Analysis}. 3rd ed. Hoboken, NJ: Wiley; 2003. }

\leavevmode\hypertarget{ref-Klein2003}{}%
\CSLLeftMargin{25. }
\CSLRightInline{Klein JP, Moeschberger ML. {Survival Analysis: Techniques for Censored and Truncated Data} {[}Internet{]}. 2nd ed. Springer-Verlag; 2003. Available from: \url{http://www.springer.com/statistics/life+sciences,+medicine+\%7B/\&\%7D+health/book/978-0-387-95399-1}}

\leavevmode\hypertarget{ref-Rothman2008h}{}%
\CSLLeftMargin{26. }
\CSLRightInline{Rothman KJ, Greenland S, Lash TL. {Modern Epidemiology}. 3rd ed. Rothman KJ, Greenland S, Lash TL, editors. Lippincott Williams {\&} Wilkins; 2008. }

\leavevmode\hypertarget{ref-Keiding1991}{}%
\CSLLeftMargin{27. }
\CSLRightInline{Keiding N. {Age-Specific Incidence and Prevalence: A Statistical Perspective}. Journal of the Royal Statistical Society Series A (Statistics in Society). 1991;154(3):371--412. }

\leavevmode\hypertarget{ref-Hawkins2005}{}%
\CSLLeftMargin{28. }
\CSLRightInline{Hawkins N, Sculpher M, Epstein D. {Cost-effectiveness analysis of treatments for chronic disease: Using R to incorporate time dependency of treatment response.} Medical Decision Making {[}Internet{]}. 2005;25(5):511--9. Available from: \url{http://www.ncbi.nlm.nih.gov/pubmed/16160207}}

\leavevmode\hypertarget{ref-Krijkamp2018}{}%
\CSLLeftMargin{29. }
\CSLRightInline{Krijkamp EM, Alarid-Escudero F, Enns EA, Jalal HJ, Hunink MGM, Pechlivanoglou P. {Microsimulation Modeling for Health Decision Sciences Using R: A Tutorial}. Medical Decision Making {[}Internet{]}. 2018 Apr;38(3):400--22. Available from: \url{http://journals.sagepub.com/doi/10.1177/0272989X18754513}}

\leavevmode\hypertarget{ref-Kuntz2001}{}%
\CSLLeftMargin{30. }
\CSLRightInline{Kuntz KM, Weinstein MC. {Modelling in economic evaluation}. In: Drummond MF, McGuire A, editors. Economic evaluation in health care: Merging theory with practice. 2nd ed. New York, NY: Oxford University Press; 2001. p. 141--71. }

\leavevmode\hypertarget{ref-Ades2004a}{}%
\CSLLeftMargin{31. }
\CSLRightInline{Ades AE, Lu G, Claxton K. {Expected value of sample information calculations in medical decision modeling.} Medical Decision Making {[}Internet{]}. 2004;24(2):207--27. Available from: \url{http://www.ncbi.nlm.nih.gov/pubmed/15090106}}

\leavevmode\hypertarget{ref-Gidwani2020}{}%
\CSLLeftMargin{32. }
\CSLRightInline{Gidwani R, Russell LB. {Estimating Transition Probabilities from Published Evidence: A Tutorial for Decision Modelers}. PharmacoEconomics {[}Internet{]}. 2020;38(11):1153--64. Available from: \url{https://doi.org/10.1007/s40273-020-00937-z}}

\leavevmode\hypertarget{ref-Rodriguez2021}{}%
\CSLLeftMargin{33. }
\CSLRightInline{Rodriguez PJ, Ward ZJ, Long MW, Austin SB, Wright DR. {Applied Methods for Estimating Transition Probabilities from Electronic Health Record Data}. Medical Decision Making {[}Internet{]}. 2021;41(2):143--52. Available from: \url{https://journals.sagepub.com/doi/abs/10.1177/0272989X20985752?casa_token=SIhw-ApH8ogAAAAA:lxalsZ3tBihW8nVpK1WhSjkuVNR3Z7kKUe44gC5UJk4Rs8KfYoAQ6Y9efV8oVvR1GeLeTDwWdRXKtWk}}

\leavevmode\hypertarget{ref-Welton2005}{}%
\CSLLeftMargin{34. }
\CSLRightInline{Welton NJ, Ades AE. {Estimation of markov chain transition probabilities and rates from fully and partially observed data: uncertainty propagation, evidence synthesis, and model calibration.} Medical Decision Making {[}Internet{]}. 2005;25(6):633--45. Available from: \url{http://www.ncbi.nlm.nih.gov/pubmed/16282214}}

\leavevmode\hypertarget{ref-Filipovic-Pierucci2017}{}%
\CSLLeftMargin{35. }
\CSLRightInline{Filipović-Pierucci A, Zarca K, Durand-Zaleski I. {Markov Models for Health Economic Evaluation: The R Package heemod}. arXiv:170203252v1 {[}Internet{]}. 2017;April:30. Available from: \url{http://arxiv.org/abs/1702.03252}}

\leavevmode\hypertarget{ref-Incerti2021}{}%
\CSLLeftMargin{36. }
\CSLRightInline{Incerti D, Jansen JP. {hesim: Health Economic Simulation Modeling and Decision Analysis}. arXiv:210209437v2 {[}Internet{]}. 2021 Feb;March:38. Available from: \url{http://arxiv.org/abs/2102.09437}}

\end{CSLReferences}
\newpage

\begin{landscape}

\section*{Supplementary Material}

\hypertarget{cohort-tutorial-model-components}{%
\subsection*{Cohort tutorial model
components}\label{cohort-tutorial-model-components}}

This table contains an overview of the key model components used in the
code for the Sick-Sicker example from the
\href{http://darthworkgroup.com/}{DARTH} manuscript:
\href{https://arxiv.org/abs/2108.13552}{``A Tutorial on Time-Dependent
Cohort State-Transition Models in R''}. The first column gives the
mathematical notation for some of the model components that are used in
the equations in the manuscript. The second column gives a description
of the model component with the R name in the third column. The forth
gives the data structure, e.g.~scalar, list, vector, matrix etc, with
the according dimensions of this data structure in the fifth column. The
final column indicated the type of data that is stored in the data
structure, e.g.~numeric (5.2,6.3,7.4), category (A,B,C), integer
(5,6,7), logical (TRUE, FALSE).

\begin{longtable}[]{@{}
  >{\raggedright\arraybackslash}p{(\columnwidth - 10\tabcolsep) * \real{0.07}}
  >{\raggedright\arraybackslash}p{(\columnwidth - 10\tabcolsep) * \real{0.35}}
  >{\raggedright\arraybackslash}p{(\columnwidth - 10\tabcolsep) * \real{0.14}}
  >{\raggedright\arraybackslash}p{(\columnwidth - 10\tabcolsep) * \real{0.14}}
  >{\raggedright\arraybackslash}p{(\columnwidth - 10\tabcolsep) * \real{0.17}}
  >{\raggedright\arraybackslash}p{(\columnwidth - 10\tabcolsep) * \real{0.12}}@{}}
\toprule
Element & Description & R name & Data structure & Dimensions & Data
type \\
\midrule
\endhead
\(n_t\) & Time horizon & \texttt{n\_cycles} & scalar & & numeric \\
& Cycle length & \texttt{cycle\_length} & scalar & & numeric \\
\(v_s\) & Names of the health states & \texttt{v\_names\_states} &
vector & \texttt{n\_states} x 1 & character \\
\(n_s\) & Number of health states & \texttt{n\_states} & scalar & &
numeric \\
\(n_{S_{tunnels}}\) & Number of health states with tunnels &
\texttt{n\_states\_tunnels} scalar & & numeric & \\
\(v_{str}\) & Names of the strategies & \texttt{v\_names\_str} & scalar
& & character \\
\(n_{str}\) & Number of strategies & \texttt{n\_str} & scalar & &
character \\
\(d_c\) & Discount rate for costs & \texttt{d\_c} & scalar & &
numeric \\
\(d_e\) & Discount rate for effects & \texttt{d\_e} & scalar & &
numeric \\
\(\mathbf{d_c}\) & Discount weights vector for costs & \texttt{v\_dwc} &
vector & (\texttt{n\_t} x 1 ) + 1 & numeric \\
\(\mathbf{d_e}\) & Discount weights vector for effects & \texttt{v\_dwe}
& vector & (\texttt{n\_t} x 1 ) + 1 & numeric \\
& Sequence of cycle numbers & \texttt{v\_cycles} & vector &
(\texttt{n\_t} x 1 ) + 1 & numeric \\
\(\mathbf{wcc}\) & Within-cycle correction weights & \texttt{v\_wcc} &
vector & (\texttt{n\_t} x 1 ) + 1 & numeric \\
\(age_{_0}\) & Age at baseline & \texttt{n\_age\_init} & scalar & &
numeric \\
\(age\) & Maximum age of follow up & \texttt{n\_age\_max} & scalar & &
numeric \\
\(M\) & Cohort trace matrix & \texttt{m\_M} & matrix & (\texttt{n\_t} +
1) x \texttt{n\_states} & numeric \\
\(M_{tunnels}\) & Aggregated Cohort trace for state-dependency &
\texttt{m\_M\_tunnels} & matrix & (\texttt{n\_t} + 1) x
\texttt{n\_states} & numeric \\
& List of the cohort trace matrix for all strategies & \texttt{l\_m\_M}
& list & & numeric \\
\(m_0\) & Initial state vector & \texttt{v\_m\_init} & vector & 1 x
\texttt{n\_states} & numeric \\
\(m_t\) & State vector in cycle \(t\) & \texttt{v\_mt} & vector & 1 x
\texttt{n\_states} & numeric \\
& & & & & \\
& \textbf{Life table input} & & & & \\
& State vector in cycle \(t\) & \texttt{lt\_usa\_2005} & list & &
numeric \\
& Vector of age-specific mortality rates & \texttt{v\_r\_mort\_by\_age}
& vector & & numeric \\
& & & & & \\
& \textbf{Transition probabilities} & & & & \\
\(p_{[H,S1]}\) & From Healthy to Sick conditional on surviving &
\texttt{p\_HS1} & scalar & & numeric \\
\(p_{[S1,H]}\) & From Sick to Healthy conditional on surviving &
\texttt{p\_S1H} & scalar & & numeric \\
\(p_{[S1,S2]}\) & From Sick to Sicker conditional on surviving &
\texttt{p\_S1S2} & scalar & & numeric \\
\(r_{[H,D]}\) & Constant rate of dying when Healthy (all-cause mortality
rate) & \texttt{r\_HD} & scalar & & numeric \\
\(r_{[S1,S2]}\) & Constant rate of becoming Sicker when Sick &
\texttt{r\_S1S2} & scalar & & numeric \\
\(r_{[S1,S2]_{trtB}}\) & Constant rate of becoming Sicker when Sick for
Treatment B & \texttt{r\_S1S2\_trtB} & scalar & & numeric \\
\(hr_{[S1,H]}\) & Hazard ratio of death in Sick vs Healthy &
\texttt{hr\_S1} & scalar & & numeric \\
\(hr_{[S2,H]}\) & Hazard ratio of death in Sicker vs Healthy &
\texttt{hr\_S2} & scalar & & numeric \\
\(hr_{[S1,S2]_{trtB}}\) & Hazard ratio of becoming Sicker when Sick
under Treatment B & \texttt{hr\_S1S2\_trtB} & scalar & & numeric \\
\(p_{[S1,S2]_{trtB}}\) & probability to become Sicker when Sick under
Treatment B conditional on surviving & \texttt{p\_S1S2\_trtB} & scalar &
& numeric \\
& & & & & \\
& \textbf{Weibull parameters for transition probability of becoming
Sicker when Sick conditional on surviving} & & & & \\
\(\lambda\) & scale of the Weibull hazard function &
\texttt{r\_S1S2\_scale} & scalar & & numeric \\
\(\gamma\) & shape of the Weibull hazard function &
\texttt{r\_S1S2\_shape} & scalar & & numeric \\
& & & & & \\
& \textbf{Simulation-time dependent mortality} & & & & \\
\(r_{[H,D,t]}\) & Age-specific background mortality rates &
\texttt{v\_r\_HDage} & vector & \texttt{n\_t} x 1 & numeric \\
\(r_{[S1,D,t]}\) & Age-specific mortality rates in the Sick state &
\texttt{v\_r\_S1Dage} & vector & \texttt{n\_t} x 1 & numeric \\
\(r_{[S2,D,t]}\) & Age-specific mortality rates in the Sicker state &
\texttt{v\_r\_S2Dage} & vector & \texttt{n\_t} x 1 & numeric \\
\(p_{[H,D,t]}\) & Age-specific mortality risk in the Healthy state &
\texttt{v\_p\_HDage} & vector & \texttt{n\_t} x 1 & numeric \\
\(p_{[S1,D,t]}\) & Age-specific mortality rates in the Sick state &
\texttt{v\_p\_S1Dage} & vector & \texttt{n\_t} x 1 & numeric \\
\(p_{[S2,D,t]}\) & Age-specific mortality rates in the Sicker state &
\texttt{v\_p\_S2Dage} & vector & \texttt{n\_t} x 1 & numeric \\
\(p_{[S1,S2, t]}\) & Time-dependent transition probabilities from sick
to sicker & \texttt{v\_p\_S1S2\_tunnels} & vector & \texttt{n\_t} x 1 &
numeric \\
\(r_{[S1,S2, t]}\) & State-residence-dependent transition rate of
becoming Sicker when Sick & \texttt{v\_r\_S1S2\_tunnels} & vector &
\texttt{n\_t} x 1 & numeric \\
& & & & & \\
& \textbf{Annual costs} & & & & \\
& Healthy individuals & \texttt{c\_H} & scalar & & numeric \\
& Sick individuals in Sick & \texttt{c\_S1} & scalar & & numeric \\
& Sick individuals in Sicker & \texttt{c\_S2} & scalar & & numeric \\
& Dead individuals & \texttt{c\_D} & scalar & & numeric \\
& Additional costs Treatment A & \texttt{c\_trtA} & scalar & &
numeric \\
& Additional costs Treatment B & \texttt{c\_trtB} & scalar & &
numeric \\
& Vector of state costs for a strategy & \texttt{v\_c\_str} & vector & 1
x \texttt{n\_states} & numeric \\
& List that stores the vectors of state costs for each strategy &
\texttt{l\_c} & List & & numeric \\
& & & & & \\
& \textbf{Utility weights} & & & & \\
& Healthy individuals & \texttt{u\_H} & scalar & & numeric \\
& Sick individuals in Sick & \texttt{u\_S1} & scalar & & numeric \\
& Sick individuals in Sicker & \texttt{u\_S2} & scalar & & numeric \\
& Dead individuals & \texttt{u\_D} & scalar & & numeric \\
& Treated with Treatment A & \texttt{u\_trtA} & scalar & & numeric \\
& Vector of state utilities for a strategy & \texttt{v\_u\_str} & vector
& 1 x \texttt{n\_states} & numeric \\
& Vector of S1 utilities when including state-residency for a strategy
SoC for & \texttt{v\_u\_S1\_SoC} & vector & 1 x \texttt{n\_tunnel\_size}
& numeric \\
& List that stores the vectors of state utilities for each strategy &
\texttt{l\_u} & List & & numeric \\
& & & & & \\
& \textbf{Transition weights} & & & & \\
& Utility decrement of healthy individuals when transitioning to S1 &
\texttt{du\_HS1} & scalar & & numeric \\
& Cost of healthy individuals when transitioning to S1 &
\texttt{ic\_HS1} & scalar & & numeric \\
& Cost of dying & \texttt{ic\_D} & scalar & & numeric \\
& & & & & \\
& \textbf{Tunnel state structures} & & & & \\
& number of tunnel states & \texttt{n\_tunnel\_size} & scalar & &
numeric \\
& vector with cycles for tunnels states & \texttt{v\_cycles\_tunnel} &
vector & 1 x \texttt{n\_tunnel\_size} & numeric \\
& tunnel names of the Sick state & \texttt{v\_Sick\_tunnel} & vector & 1
x \texttt{n\_states} & numeric \\
& state names including tunnel states &
\texttt{v\_names\_states\_tunnels} & vector & 1 x
\texttt{n\_states\_tunnels} & character \\
& number of states including tunnel states & \texttt{n\_states\_tunnels}
& scalar & & numeric \\
& Initial state vector for the model with tunnels &
\texttt{v\_m\_init\_tunnels} & vector & 1 x \texttt{n\_states\_tunnels}
& numeric \\
& & & & & \\
\(\mathbf{P}\) & Time-dependent transition probability array &
\texttt{a\_P} & array & \texttt{n\_states} x \texttt{n\_states} x
\texttt{n\_t} & numeric \\
\(\mathbf{P}_{tunnels}\) & Transition probability array for the model
with tunnels & \texttt{a\_P\_tunnels} & array &
\texttt{n\_states\_tunnels} x \texttt{n\_states\_tunnels} x
\texttt{n\_t} & numeric \\
\(\mathbf{A}\) & Transition dynamics array & \texttt{a\_A} & array &
\texttt{n\_states} x \texttt{n\_states} x (\texttt{n\_t} + 1) &
numeric \\
& List of the transition dynamics arrays for all strategies &
\texttt{l\_m\_A} & list & & numeric \\
\(\mathbf{R_u}\) & Transition rewards for effects & \texttt{a\_R\_u} &
array & \texttt{n\_states} x \texttt{n\_states} x (\texttt{n\_t} + 1) &
numeric \\
\(\mathbf{R_c}\) & Transition rewards for costs & \texttt{a\_R\_c} &
array & \texttt{n\_states} x \texttt{n\_states} x (\texttt{n\_t} + 1) &
numeric \\
\(\mathbf{Y_u}\) & Expected effects per states per cycle &
\texttt{a\_Y\_u} & array & \texttt{n\_states} x \texttt{n\_states} x
(\texttt{n\_t} + 1) & numeric \\
\(\mathbf{Y_c}\) & Expected costs per state per cycle & \texttt{a\_Y\_c}
& array & \texttt{n\_states} x \texttt{n\_states} x (\texttt{n\_t} + 1)
& numeric \\
& & & & & \\
& Expected QALYs per cycle under a strategy & \texttt{v\_qaly\_str} &
vector & 1 x (\texttt{n\_t} + 1) & numeric \\
& Expected costs per cycle under a strategy & \texttt{v\_cost\_str} &
vector & 1 x (\texttt{n\_t} + 1) & numeric \\
& Vector of expected discounted QALYs for each strategy &
\texttt{v\_tot\_qaly} & vector & 1 x \texttt{n\_states} & numeric \\
& Vector of expected discounted costs for each strategy &
\texttt{v\_tot\_cost} & vector & 1 x \texttt{n\_states} & numeric \\
& Summary matrix with costs and QALYS per strategy &
\texttt{m\_outcomes} & table & \texttt{n\_states} x 2 & \\
& Summary of the model outcomes & \texttt{df\_cea} & data frame & & \\
& Summary of the model outcomes & \texttt{table\_cea} & table & & \\
& Input parameters values of the model for the cost-effectiveness
analysis & \texttt{df\_psa} & data frame & & \\
& *NOTE: these structures can have \texttt{\_strX} to indicate the
strategy of interest & & & & \\
& & & & & \\
& \textbf{Probabilistic analysis structures} & & & & \\
& Number of PSA iterations & \texttt{n\_sim} & scalar & & numeric \\
& List that stores all the values of the input parameters &
\texttt{l\_params\_all} & list & & numeric \\
& Data frame with the parameter values for each PSA iteration &
\texttt{df\_psa\_input} & data frame & & numeric \\
& Vector with the names of all the input parameters &
\texttt{v\_names\_params} & vector & & character \\
& List with the model outcomes of the PSA for all strategies &
\texttt{l\_psa} & list & & numeric \\
& Vector with a sequence of relevant willingness-to-pay values &
\texttt{v\_wtp} & vector & & numeric \\
& Data frame to store expected costs and effects for each strategy from
the PSA & \texttt{df\_out\_ce\_psa} & data frame & & numeric \\
& Data frame to store incremental cost-effectiveness ratios (ICERs) from
the PSA & \texttt{df\_cea\_psa} & data frame & & numeric \\
& For more details about the PSA structures read \texttt{dampack}'s
vignettes & & & & \\
\bottomrule
\end{longtable}

\end{landscape}
\end{document}